\documentclass[%
 reprint,
 prapplied,
 groupedaddress,
superscriptaddress,
 amsmath,amssymb,
 aps,
]{revtex4-2}
\usepackage{graphicx}
\usepackage{dcolumn}
\usepackage{bm}
\usepackage{siunitx}
\usepackage{physics}
\usepackage[version=4]{mhchem}
\usepackage{txfonts}
\usepackage{hyperref}
\usepackage{hyperref}
\hypersetup{
    colorlinks=true,
    linkcolor=blue,
    filecolor=magenta,      
    urlcolor=blue,
    pdftitle={Widefield AC Susceptibility},
    pdfauthor={Dasika Shishir},
    }

\begin{document}

\preprint{}

\title{Observation of Ultra-low AC Susceptibility in Micro-magnets Using Quantum Diamond Microscope}
\begin{abstract}
AC susceptometry, unlike static susceptometry, offers a deeper insight into magnetic materials. By employing AC susceptibility measurements, one can glean into crucial details regarding magnetic dynamics. Nevertheless, traditional AC susceptometers are constrained to measuring changes in magnetic moments within the range of a few nano-joules per tesla. Additionally, their spatial resolution is severely limited, confining their application to bulk samples only. In this study, we introduce the utilization of a Nitrogen Vacancy (NV) center-based quantum diamond microscope for mapping the AC susceptibility of micron-scale ferromagnetic specimens. By employing coherent pulse sequences, we extract both  magnitude and the phase of the field  from samples within a field of view spanning  70 micro-meters while achieving a resolution of 1 micro-meter. Furthermore, we quantify changes in dipole moment on the order of a  femto-joules per tesla induced by excitations at frequencies reaching several hundred kilohertz.


\end{abstract}
    
\author{Dasika Shishir}
\affiliation{Department of Electrical Engineering, Indian Institute of Technology Bombay, Mumbai, Maharashtra, India}
\email{dshishir@iitb.ac.in}
\author{Matthew L. Markham}
\affiliation{Element Six, Harwell OX11 0QR, United Kingdom}
\author{Kasturi Saha}
\email{kasturis@ee.iitb.ac.in}
\affiliation{Department of Electrical Engineering, Indian Institute of Technology Bombay, Mumbai, Maharashtra, India}
\affiliation{Center of Excellence in Quantum Information, Computing Science and Technology, Indian Institute of Technology Bombay, Powai, Mumbai--400076, India}
\affiliation{Center of Excellence in Semiconductor Technologies (SemiX), Indian Institute of Technology Bombay, Powai, Mumbai--400076, India}

\maketitle


\section{Introduction}

Micro and nano-scale magnetic materials play crucial role across a spectrum of scientific, 
industrial, and research domains. Recent advancements have underscored their significance in 
diverse applications, such as utilizing magnetic states in micro-magnets for memory storage\,\cite{Ramasubramanian2020,doi:10.1126/science.289.5481.930}, employing magnetic nanoparticle-based 
biomedicine\,\cite{pharmaceutics13070943}, exploring biosensing 
capabilities\,\cite{https://doi.org/10.1002/elan.200603785}, investigating exotic quantum phases 
like the quantum spin Hall effect,  spin liquids\,\cite{Burch2018}, and pioneering 
spin-based computing\,\cite{HIROHATA2020166711}. Central to understanding these materials is the magnetic 
susceptibility, a pivotal parameter characterizing their magnetic properties. Particularly, the 
measurement of AC susceptibility\,\cite{Topping_2019} offers insights into ferromagnetic dynamics, 
including domain wall velocity and vortex motion\,\cite{Topping_2019,cullity2011introduction}. However, conventional tools for assessing AC 
susceptibility are limited, typically detecting changes in magnetic moments of just a few \SI{}{\nano \joule \per \tesla}\,\cite{cullity2011introduction,10.1063/1.5046475}. To 
contextualize this quantity, consider that the magnetic moment of a saturated cast iron cube—measuring approximately \SI{11}{\um} per side, with a saturation magnetization of 
\SI{8e5}{\ampere \per \meter} is around  \SI{1}{\nano \joule \per \tesla}. Remarkably, in a thin 
film of iron with the same lateral dimensions but a thickness of about 10 nanometers, the magnetic 
moment decreases to a mere \SI{1}{\pico \joule \per \tesla}. Under unsaturated conditions, 
magnetic moments can even be of the order of \SI{1}{\femto \joule \per \tesla}.

In recent years, there has been a surge in the characterization of magnetic materials at the nanoscale, facilitated by the remarkable magnetic field sensing capabilities of negatively charged nitrogen vacancy centers in diamond (\ce{NV-} centers in diamond). This advancement has enabled the characterization of a broad range of materials, including conventional ferromagnets\,\cite{Toraille2018,10.1063/1.3337096,doi:10.1126/science.aaw4352,10.1063/5.0138301,PhysRevB.88.214408}, 2D van der Waals magnets\,\cite{Huang2023,https://doi.org/10.1002/adma.202003314,PhysRevB.109.064416,Healey2022,10.1063/5.0091931,PRXQuantum.2.030352,Robertson_2023}, and magnetic particles found in biological systems\,\cite{doi:10.1073/pnas.2112749118,doi:10.1126/sciadv.adi5300,doi:10.1073/pnas.2112664118}. Most studies in this realm have focused on measuring the static or DC magnetic susceptibility, typically achieved by gauging the stray magnetic field emitted by the material under various external magnetic field conditions. Notably, in many instances, the magnitude of the stray field from the sample is significantly smaller than the applied magnetic field \,\cite{10.1063/5.0138301,doi:10.1073/pnas.2112749118}.

In a recent study (Ref.\,\cite{PRXQuantum.2.030352}), both AC and static susceptibility 
measurements were conducted with a confocal microscope by demagnetizing the sample, elevating its temperature beyond its 
Curie point, and then assessing the stray fields. This approach 
allows  to differentiate between the field emanating from the sample and the applied field. 
Alternatively, the utilization of \ce{NV-} center based quantum widefield diamond microscopes (QDM)\,
\cite{Parashar2022,LevineTurnerKehayiasHartLangellierTrubkoGlennFuWalsworth+2019+1945+1973}, 
offers another avenue to distinguish the applied 
field from the stray field of the sample. In a QDM, the magnetic field across a wide field of view (ranging from 10 micrometers to 1 
millimeter) is mapped as in Ref.\,\cite{10.1063/5.0138301}, with regions lacking magnetic material 
providing insights into the background or applied magnetic field. By subtracting the applied 
magnetic field, information regarding the stray magnetic field from the sample can be extracted. 
However, the application of continuous-wave optically detected magnetic resonance (ODMR) to sense 
magnetic fields from materials has been limited to low frequencies, typically in the range of 
several tens of kilohertz \,\cite{PhysRevLett.106.030802}.

Here, we present a novel approach for measuring the AC susceptibility of permalloy (Py) micro-disks, each spanning micrometer dimensions, using QDM. We employ XY dynamical decoupling sequences in widefield\,\cite{PhysRevApplied.16.054014,Bucher2019} to detect small AC magnetic fields emanating from the magnetic micro-disks at high frequencies. This technique allows us to discern changes in magnetic moment induced by the applied AC fields, which typically hover around \SI{1}{\femto \joule \per \tesla}, with a spatial resolution of approximately around \SI{1}{\um}. Such precise measurements would pose significant challenges with conventional methodologies. Additionally, we delve into the implications of various non-idealities and imperfections that may influence the accuracy of these measurements.

\section{Results and Discussion}
\subsection{Implementation}
\begin{figure}[ht]
\centering
\includegraphics[width=\linewidth]{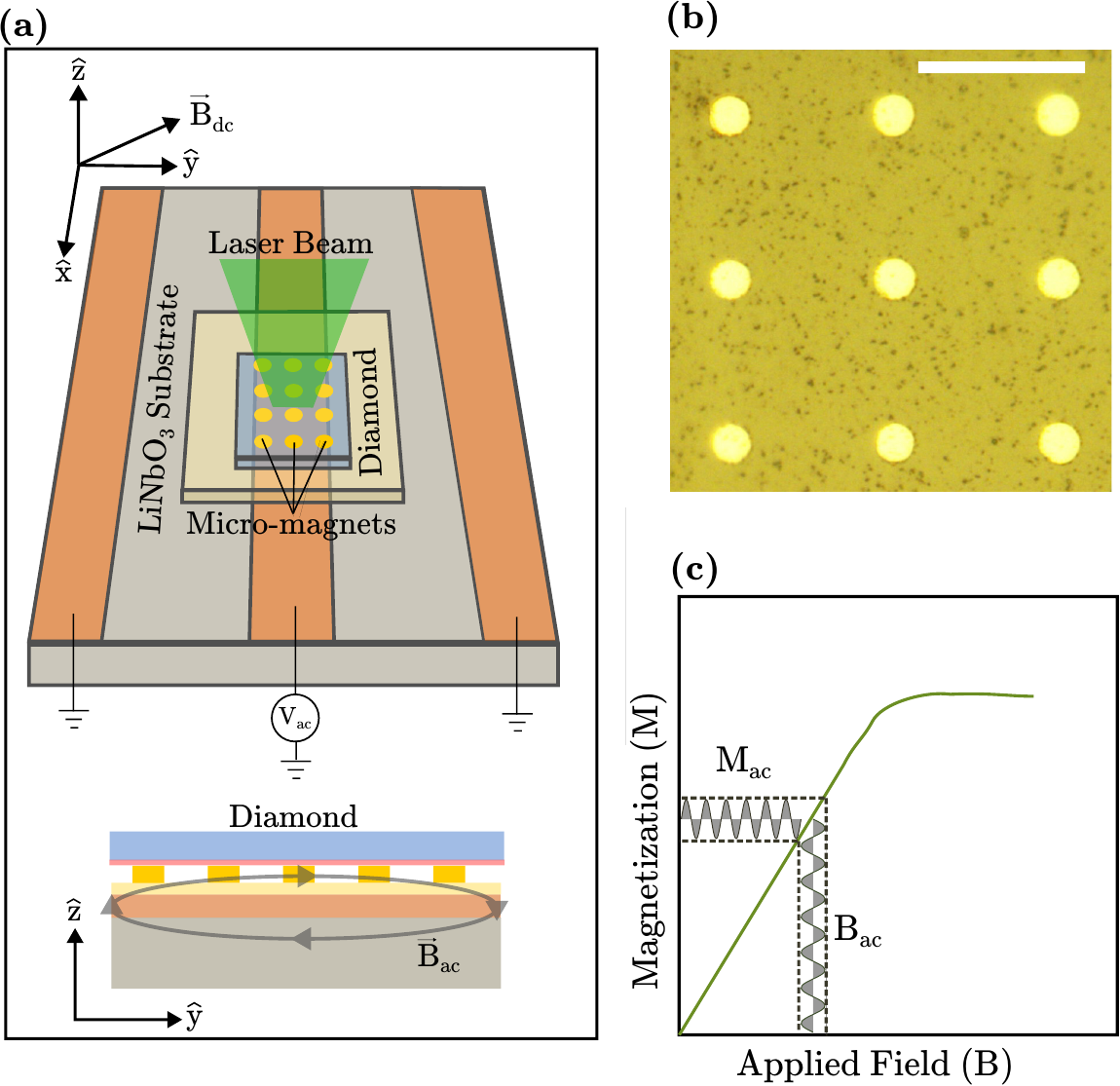}
\caption{Exeperimental platform. (a). Permalloy micro-magnets of \SI{5.8\pm 0.4}{\micro\meter} diameter 
are patterned on a \ce{LiNbO3} substrate. A permanent magnet is used to apply the static field $B_{\text{DC}}$
along a \ce{NV-} axis. A \SI{4}{\mm} wide strip-line is used to provide the AC excitation 
$B_{\text{AC}}$. The AC field generated at the center of the strip-line is mostly along the
$\vu{y}$-axis. Approximately \SI{100}{\micro \meter} wide laser beam is used to excite the 
diamond. (b). Optical microscope image of the micro-magnets. 
The scale bar is  \SI{25}{\micro\meter}. 
(c). To measure the AC susceptibility, an AC field $B_{\text{AC}}$ is provided to the 
micro-magnets,  which causes the magnetization of the micro-magnets to oscillate. The resultant oscillating
stray field from the micro-magnets is measured by the \ce{NV-} centers in diamond.}
\label{fig:schematic}
\end{figure}

Figure \ref{fig:schematic}(a) illustrates the experimental setup employed for 
measuring  the DC, and the AC susceptibility of the micro-magnets. The \ce{Py} 
micro-magnets
are deposited on a \SI{0.5}{\mm} thick Lithium Niobate (\ce{LiNbO3}) substrate.
The diameter of each micro-magnet is \SI{5.8\pm 0.4}{\um},
and the thickness is \SI{30}{\nm}. 
The center-to-center distance between the micro-magnets is \SI{25}{\um} as 
shown in Fig.\,\ref{fig:schematic}(b). 
The substrate is placed on a \SI{4}{\mm} wide copper strip, which generates the AC excitation field 
(${B}_{\text{AC}}$). The field from the copper strip is predominantly in the
plane of the micro-magnets, along $\vu{y}$-axis, as depicted in Fig.\,
\ref{fig:schematic}(a). A current of \SI{1}{\ampere} through the strip-line
generates a field of
\SI{100}{\micro\tesla} along the \ce{NV-} axis. The 
static bias field ($\va{B}_{\text{DC}}$) is established by a \ce{SmCo} permanent magnet, 
oriented along one of the crystallographic axes of the \ce{NV-} center. 
The \ce{NV-} makes  an angle 
of $\theta = 54.6^\circ$ relative to the surface normal of the diamond, in the
$\vu{y}-\vu{z}$ plane as shown in Fig.\,\ref{fig:schematic}(a). The 
static field can be adjusted within a range of \SI{0.8}{\milli \tesla} to 
\SI{5}{\milli \tesla}.

An optical excitation of \SI{0.3}{\watt} at \SI{532}{\nm} is delivered through a 
\num{0.9} numerical aperture objective onto the diamond. The lateral dimensions
of the diamond are \qtyproduct{4x4}{\mm}, and the thickness is \SI{0.25}{\mm}.
The 
diamond features a $\qty{100}$ front facet and $\langle 110 \rangle$ edge 
orientation, with a 
high isotopic purity of \SI{99.99}{\percent}, and a surface layer of 2 ppm 
\ce{^{15}N}, 
approximately \SI{1}{\um} thick. The resulting red fluorescence is directed onto 
a high-speed lock-in camera after passing through requisite optical filters. The 
overall field of view spans \qtyproduct{70 x 70}{\um} with a each pixel bieng \SI{1}{\um}, and an overall magnification of 40. Further 
details are provided in  Appendix.\,\ref{sec:exp_details}. 
   
\subsection{Widefield DC Magnetometry}
We first characterize the static properties of the micro-magnets by varying the
external static field $B_{\text{DC}}$, and measuring the dipole moment of the
micro-magnets. 
We derive the local magnetic field corresponding to each applied external 
magnetic field by examining the Zeeman splitting observed in the optically 
detected magnetic resonance (ODMR) signal from individual pixels of the 
camera. The Zeeman 
frequency encompasses contributions from both the applied external magnetic 
field, and the magnetic field generated by the micro-magnets. To differentiate 
the  external magnetic field from the field of the micro-magnets, we 
first, filter the magnetic field image with  
a Gaussian filter which has a kernel size of \num{50} pixels. Subsequently, we 
subtract the filtered  image from the original image, as outlined in 
\cite{doi:10.1073/pnas.2112749118} to obtain the field from the sample
$B_{DC}^s$. 

\begin{figure}
 \centering
 \includegraphics[width=\linewidth]{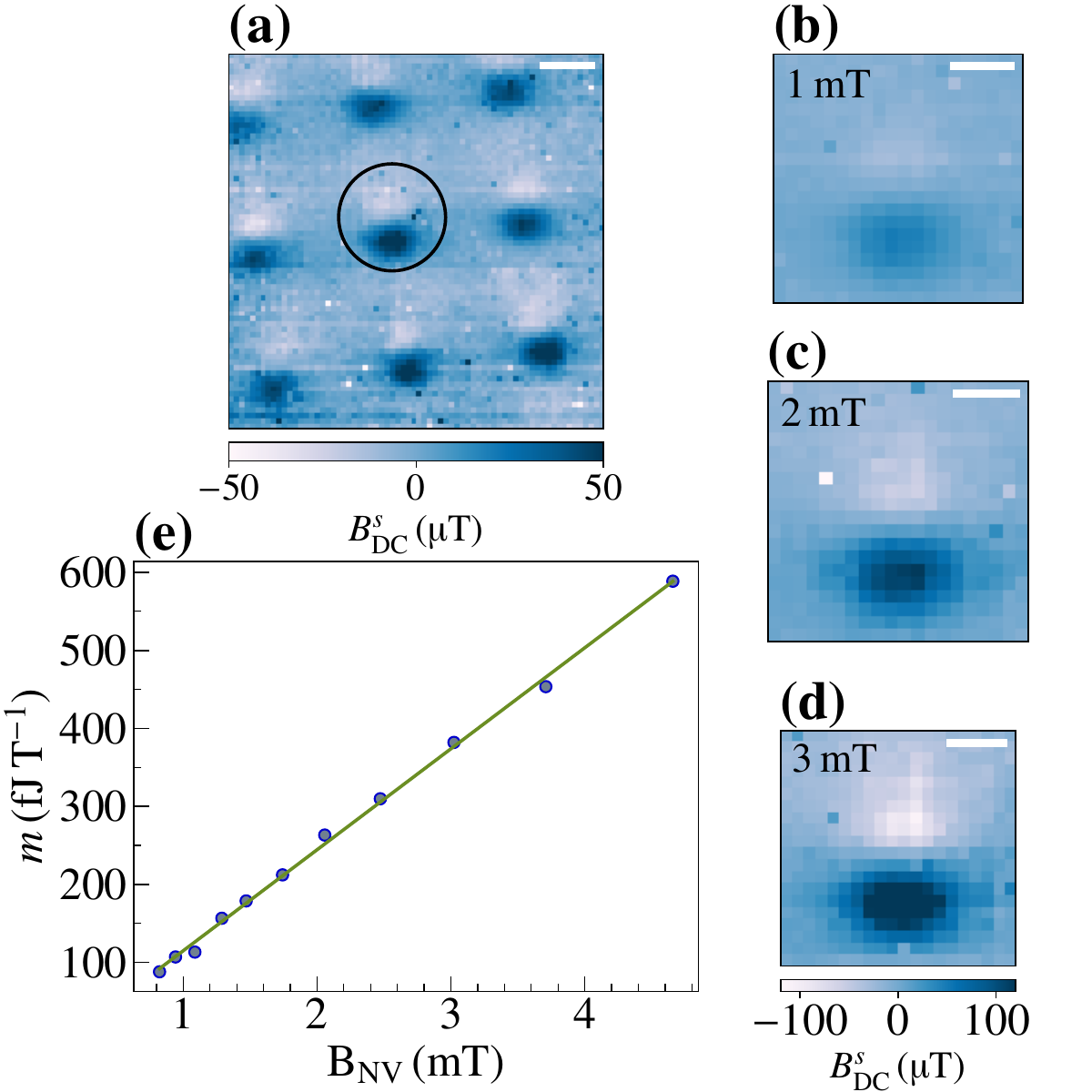}
 \caption{Static micro-magnet characterization. (a). Static magnetic field from the micro-magnets when \SI{1}
 {\milli \tesla} field is applied along the \ce{NV-} axis. The scale bar is \SI{10}{\micro \meter}. (b). Static magnetic field from  the central micro-magnet (circled in (a)) at \SI{1}{\milli \tesla}, (c) at \SI{2}{\milli \tesla}, and (d) at \SI{3}{\milli \tesla}. The scale bar for (b), (c), and (d) is \SI{5}{\micro \meter}. (e). Variation in dipole moment with applied magnetic field, fit to a straight line. }
 \label{fig:DC}
\end{figure}

In Figure \ref{fig:DC}(a), we present the static magnetic field map from the 
sample  when an external magnetic field of \SI{1}{\milli \tesla} is applied 
along the \ce{NV-} axis. The micro-magnets become magnetized, 
exhibiting a dipole moment of \SI{107 \pm 5}{\femto \joule \per \tesla}. 
Analysis of the magnetic field map allows us to estimate the offset distance 
between the \ce{NV-} layer and the micro-magnets to be approximately \SI{6\pm0.7}
{\um} (further details are provided in Appendix \ref{sec:micro_model}). As the 
sample lacks perpendicular anisotropy, magnetization predominantly aligns within 
the plane due to the relatively large lateral dimensions compared to the 
thickness\,\cite{cullity2011introduction,10.1063/1.3624900}. The volume
normalized magnetic susceptibility $\chi_V$ is defined as\,\cite{cullity2011introduction} 
\begin{equation}
 \chi_V = \frac{1}{V} \frac{\Delta m}{\Delta H_y},
\end{equation}
where $\Delta m$ denotes the change in the dipole moment for a change in magnetic 
excitation $\Delta H_y$. In our study, we adhere to SI units, where the magnetic 
dipole moment $m$ is measured in \SI{}{\joule \per \tesla} (equivalent to 
\SI{1e3}{emu}), the magnetic induction $H$ is in \SI{}{\ampere \per \meter} 
(\SI{1}{\ampere \per \meter} is equal to $4\pi\times 10^{-3}$\SI{}{Oe}), and 
the magnetic field $B = \mu_\circ H$ is in \SI{}{\tesla}. In SI units, the 
volume-normalized susceptibility is a dimensionless quantity. Given that our magnetic field 
is aligned to a single axis of the \ce{NV-} center, we solely consider the 
in-plane magnetic field for calculating the susceptibility, denoted as $H_y$.

In Figures \ref{fig:DC}(b) to \ref{fig:DC}(d), we depict the stray field emanating from the 
central micro-magnet (circled in Fig.\,\ref{fig:DC}(a)) under excitation fields 
of \SI{1}{\milli \tesla}, \SI{2}{\milli \tesla}, and \SI{3}{\milli \tesla}, 
respectively. The stray field exhibits a direct proportionality to the magnetic 
dipole moment, evident in its escalation with increasing applied field strength. 
The extraction of the micro-magnets' dipole moment follows the procedure 
detailed in Appendix \ref{sec:micro_model}.

The dipole moment demonstrates an almost linear increase at a rate of 
approximately \SI{129}{\femto \joule \per \tesla \per \milli \tesla}, as 
illustrated in Figure \ref{fig:DC}(e). Extrapolation of the linear fit to zero 
magnetic field yields a small dipole moment of 
\SI{-14}{\femto \joule \per \tesla}, indicating the absence 
of significant remnant magnetization, and suggesting a  soft magnetic nature of 
the micro-magnets. Notably, the magnetic susceptibility for the central micro-magnet 
is measured to be \num{138\pm 5} (the mean volume of each micro-magnet is 
\SI{3.17}{\micro \meter^3}), with similar values observed for the other 
micro-magnets.

\subsection{Widefield AC Magnetometry}
\begin{figure*}
 \centering
 \includegraphics[width=\linewidth]{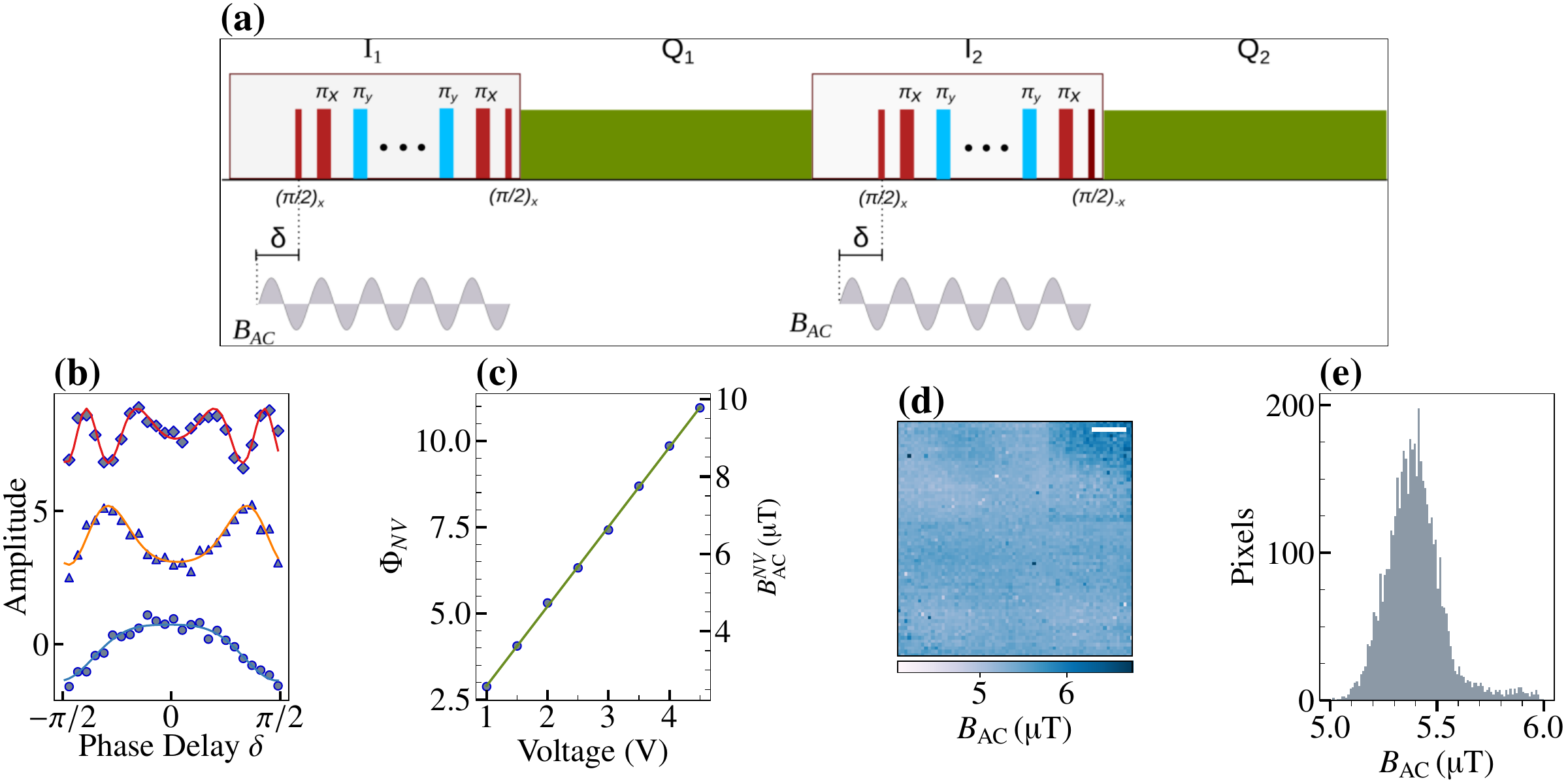}
 \caption{Widefield AC magnetometry. (a). Pulse sequence used
 for widefield AC sensing. The pulse sequence consists of 
 XY-N decoupling sequence. An external AC signal $B_{AC}$
 is provided through a current carrying strip. The initial 
 phase $\delta$ of the external AC excitation with respect to 
 the initial $\pi/2$ pulse is swept from $-\pi/2$ to $\pi/2$. (b). 
 Signals resulting from the phase shift for an applied voltage 
 of \SI{1}{\volt}, \SI{2}{\volt}, and \SI{4}{\volt} fit to 
 Eq.\,\eqref{eq:xy_signal} at a random pixel. The pulse 
 sequence used is the XY-4 sequence, and the AC signal frequency is \SI{200}{\kilo\hertz}. (c). The phase 
 acquired by the \ce{NV-} center after the end of the 
 sequence for various applied voltage fit to a straight line. (d). Widefield AC magnetic field map in the 
 absence of micro-magnet sample. The scale-bar is \SI{10}{\um}. (e). Histogram of the 
 detected AC magnetic field at an applied voltage of \SI{3}{\volt}. The standard deviation across the pixels is
 \SI{120}{\nano\tesla}.}
 \label{fig:calib}
\end{figure*}

AC susceptometry on a magnetic sample is performed as shown in Fig.\,\ref{fig:schematic}(c). 
An AC magnetic field of small amplitude is applied on top of a static magnetic field. The AC 
magnetic field results in the oscillation of sample's magnetization. The resultant AC stray field 
is measured by the \ce{NV-} centers. 
For performing widefield AC magnetometry, we use the 
XY-N dynamical decoupling 
sequences\,\cite{RevModPhys.89.035002,PRXQuantum.2.030352,PhysRevApplied.16.054014} (DD sequence). The exact pulse sequence is shown in
Fig.\,\ref{fig:calib}(a). The 
pulse sequence consists of a train of $\pi$ pulses sandwiched 
between two $\pi/2$ pulses with alternating $x$, $y$ phases. 
The train of $\pi$ pulses 
rectifies the applied AC signal which results in a net
phase $\Phi_{NV}$ acquired by the \ce{NV-} centers. The spacing $\tau$ 
between the $\pi$ pulses is equal to half the 
period of the signal that is to  be sensed, i.e. $\tau = 1/2f_{ac}$. 
The image acquisition is performed by Heliotos Helicam C3 lockin 
camera. Each demodulation cycle is divided into four parts
namely - $\text{I}_1$, $\text{Q}_1$, $\text{I}_2$, and 
$\text{Q}_2$. After every demodulation cycle, the camera outputs 
two frames : $\text{I}_2 - \text{I}_1$, and 
$\text{Q}_2 - \text{Q}_1$. In our experiment, the demodulation
rate is \SI{8}{\kilo \hertz}. Hence, $\text{I}_1$, $\text{Q}_1$, 
$\text{I}_2$, and $\text{Q}_2$ are each 
\SI{31.25}{\micro \second} long. We put the pulse sequence
in $\text{I}_1$, $\text{I}_2$, and the laser readout, and 
the polarization  in  $\text{Q}_1$, and 
$\text{Q}_2$. The pulse sequence in $\text{I}_1$, and 
$\text{I}_2$ are essentially the same, but with  
$+x$, and $-x$ final  $\pi/2$ pulse respectively. Doing so, enhances the
contrast of the signal\,\cite{PhysRevApplied.15.044020}.  
To sense the response of the micro-magnets
to an excitation frequency $f_{ac}$,
we apply an AC excitation field  whose component along the \ce{NV-} axis is given by
\begin{equation}
b_{\text{AC}}^a(t) = B_{\text{AC}}^a \cos (2\pi f_{ac} t + \delta),
    \label{eq:Bac}
\end{equation}
where, $f_{ac}$ is the excitation frequency, $\delta$ is the phase of the
signal with respect to the initial $\pi/2$ pulse in the XY-N sequence. Note that the actual 
field is along the $\vu{y}$ axis, and the component of the field along the \ce{NV-} axis is
$\sqrt{2/3}$ times of it. The sample responds to the applied AC field with a field given by
\begin{equation}
    b_{\text{AC}}^s(t) = B_{\text{AC}}^s \cos (2\pi f_{ac} t + \delta),
    \label{eq:Bsamp}
\end{equation}
where we assume that there is no phase shift in the sample signal from the applied signal
(see Appendix.\,\ref{sec:sensitivity} for a more detailed analysis).  
In this case, 
the net phase acquired, $\Phi_{\text{NV}}$, by the \ce{NV-} centers due to the DD sequence 
is given by\,\cite{RevModPhys.89.035002,PRXQuantum.2.030352,Mizuno2020}
\begin{equation}
   \Phi_{NV} =  4\gamma_e N_p \tau B_{\text{AC}} \cos \delta = \kappa B_{\text{AC}} \cos \delta, 
   \label{eq:phase}
\end{equation}
where $N_p$ is the total number of $\pi$ pulses, and $\gamma_e$ is the
electron gyro-magnetic ratio equal to 
\SI{28.024}{\kilo \hertz \per \micro \tesla}, and $B_{\text{AC}} = B_{\text{AC}}^a + B_{\text{AC}}^s$.  The phase is converted to 
a fluorescence signal by the final $\pi/2$ pulse. The fluorescence signal 
is given by\,\cite{RevModPhys.89.035002,PRXQuantum.2.030352,Mizuno2020}
\begin{equation}
 S_{\text{XY}} = C_\circ  + C\cos \left( \Phi_{\text{NV}} \cos \left(\delta \right) \right),
 \label{eq:xy_signal}
\end{equation}
where $C$ is the contrast of the signal, and $C_\circ$ is the baseline of the signal. 
Note that we are interested in extracting both the phase and the amplitude 
of the AC magnetic field. From Eq.\,\eqref{eq:xy_signal}, the sensitivity to the
phase and the amplitude of the AC signal is given by
\begin{align}
    \pdv{S_{XY}}{B_{AC}} &= -C \cdot \kappa \cdot \sin \left ( \kappa  B_{\text{AC}} \cos \delta \right) \cos \delta \label{eq:amp_sensitivity} \\
     \pdv{S_{XY}}{\delta} &= C B_{\text{AC}} \cdot \kappa \cdot \sin \left ( \kappa  B_{\text{AC}} \cos \delta \right) \sin \delta \label{eq:phase_sensitivity}. 
\end{align}
From Eq.\,\eqref{eq:amp_sensitivity}, it can be seen that the maximum sensitivity for 
detecting amplitude of the AC signal is achieved when the phase 
$\delta = 0$. However, at $\delta=0$, the sensitivity to detect any
change in phase  is zero. From Eq.\,\eqref{eq:phase_sensitivity}, 
a similar statement can be made about the sensitivity to detect the phase, which
is maximum at $\delta=\pi/2$, at which the amplitude sensitivity is zero. 
Hence, we sweep the phase of the excitation signal of Eq.\,\eqref{eq:Bac}
from $\delta=0$ to $\delta=\pi$ as Eq.\,\eqref{eq:xy_signal} is periodic with $\pi$. 

Initially, we examine the scenario where the micro-magnet sample is not present. In this instance, 
we utilize a \ce{LiNbO3} substrate devoid of any micro-magnets, maintaining the same geometry as 
depicted in Fig.\,\ref{fig:schematic}(a). To ascertain the durations of the $\pi$, and $\pi/2$ 
pulses, as well as the uniformity of the microwave drive strength, we conduct the widefield Rabi 
experiment. Due to the proximity of a neighboring hyperfine from a \ce{^{15}N} nucleus, we observe 
a composite of two sinusoidal waves. The data from a single
pixel is fit to a sum of two exponentially decaying cosine 
functions\,\cite{10.1119/5.0075519} 
\begin{equation}
 A_1 \cos (\omega_1 t)e^{\left(-t/T_1\right)} + A_2 \cos (\omega_2 t)e^{\left(-t/T_2\right)},
\end{equation}
where $\omega_1$, and $\omega_2$ are the Rabi frequencies. 
If, we assume, $\omega_1$ is the slower frequency, 
corresponding to the resonant frequency, the faster frequency
$\omega_2$ is given by
\begin{equation}
 \omega_2 = \sqrt{\omega_1^2 + \Delta^2},
 \label{eq:general_rabi}
\end{equation}
where $\Delta$ is the de-tuning from the resonance. In our
experiment the mean value of $\omega_1$ over all the pixels is 
\SI{2\pi \times 2.7}{\mega \radian\per \second}, and the mean value
of $\omega_2$ over all the pixels is \SI{2\pi \times 4}{\mega 
\radian \per \second}. Using Eq.\,\eqref{eq:general_rabi}, we 
find that $\Delta$ is equal to \SI{2\pi\times 3}{\mega 
\radian \per \second}, which is approximately equal to the 
hyper-fine 
frequency of the \ce{^{15}N} spin.  In the absence of a sample, the variation 
$\pi$ pulse times is less than \SI{6}{\nano \second}, indicating a high degree of uniformity. 
In Fig.\,\ref{fig:calib}(b), the signal obtained from a 
random pixel, for various applied voltages is shown at 
a  frequency of \SI{200}{\kilo \hertz} using a
XY-4 ($(\pi/2)_x-\tau/2-\pi_x-\tau-\pi_y-\tau-\pi_x-\tau-\pi_y-\tau/2-(\pi/2)_x$) decoupling sequence. The signals
follow the functional relationship given in 
Eq.\,\eqref{eq:xy_signal}. By fitting the equation, we 
obtain a linear relationship between the applied voltage and 
the phase $\Phi_{NV}$, acquired by the \ce{NV-} centers 
as shown in Fig.\,\ref{fig:calib}(c). The applied AC field
is proportional to the phase acquired by the \ce{NV-} centers
$\Phi_{NV}$ as given Eq.\,\eqref{eq:phase}. The resulting
widefield image is shown in Fig.\,\ref{fig:calib}(d). The 
AC field histogram across various pixels is shown in 
Fig.\,\ref{fig:calib}(e). The standard deviation of the 
sensed AC magnetic field is \SI{120}{\nano \tesla}.

\begin{figure*}
 \centering
 \includegraphics[width=\linewidth]{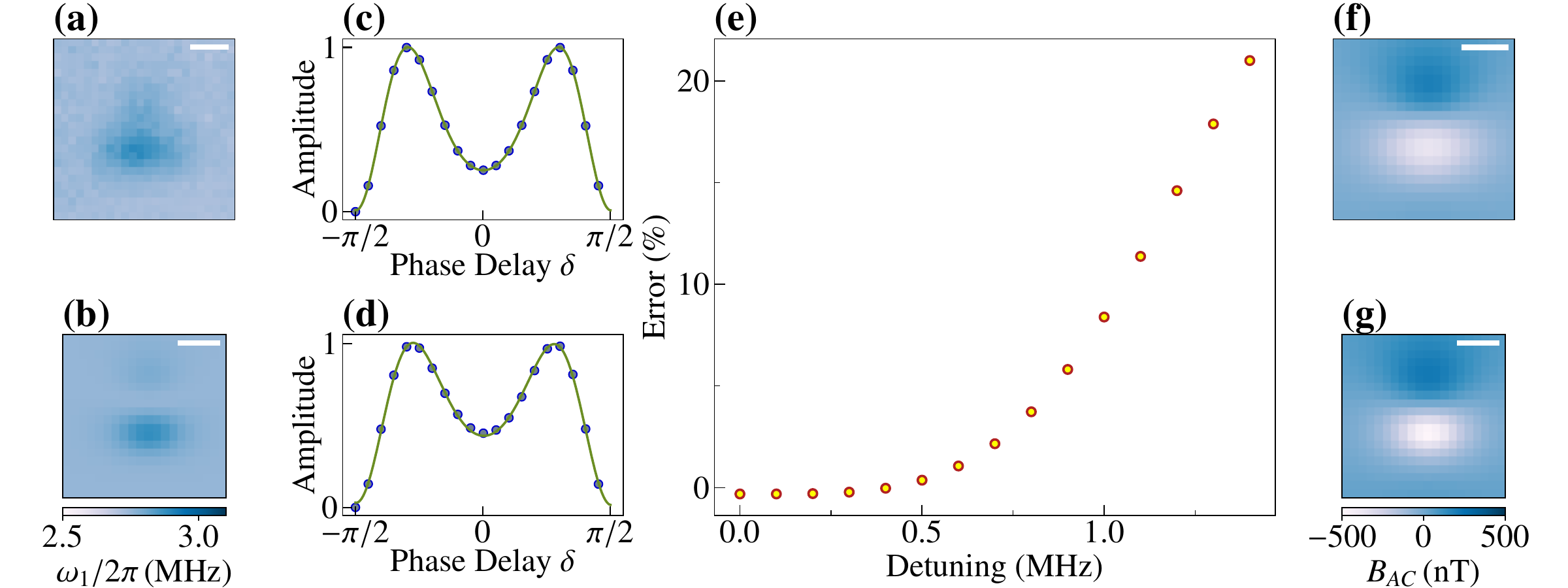}
 \caption{Effects of inhomogeneity. (a). Experimentally 
 obtained variation of Rabi frequency around a micro-magnet. The Rabi 
 frequency away from the micro-magnet is \SI{2.7}{\mega 
 \hertz}. At the location of maximum field from the micro-magnet, 
 the Rabi frequency is \SI{2.85}{\mega \hertz}. (b). 
 Simulated Rabi frequency for a micro-magnet. (c). 
 Simulated XY-8 signal for an applied AC field of \SI{3.5}
 {\micro \tesla} without any de-tuning. (d). Simulated XY-8 signal 
 for an applied AC field of \SI{3.5}{\micro \tesla} for a 
 de-tuning of \SI{0.9}{\mega\hertz}. Note the resulting signal 
 sensed AC signal is less than the applied AC signal in this
 case. (e). Percentage error in the sensed AC signal for 
 various de-tunings. The base Rabi frequency is assumed to
 be \SI{2.7}{\mega\hertz}. (f). Simulated AC field from 
 a micro-magnet in the absence of in-homogeneities. (g). Simulated AC field from 
 a micro-magnet in the presence of in-homogeneities. The scale-bar in all the figures is \SI{5}{\um}.}
 \label{fig:inhomo}
\end{figure*}

A significant challenge encountered when sensing AC fields from magnetic 
materials using dynamical decoupling sequences is the impact of DC 
inhomogeneities \,\cite{doi:10.1073/pnas.2112749118,s18041290,PhysRevA.99.012110}. The static 
magnetic field emanating from these materials can induce de-tuning 
effects on the applied $\pi$ and $\pi/2$ pulses. Since a de-tuned microwave 
signal fails to induce a complete transition from $\ket{0}$ to $\ket{1}$, errors 
may arise in the sensed field. To assess the errors stemming from these 
inhomogeneities, we initially simulate their effects. In this work, we measure
the initial AC susceptibility, when an external static magnetic field of \SI{0.8}{\milli \tesla} is applied along the \ce{NV-} axis. In Fig.\,\ref{fig:inhomo}(a), we present the 
experimentally obtained Rabi frequency map around the 
central micro-magnet under an applied static magnetic field of \SI{0.8}{\milli \tesla} along the
\ce{NV-} axis. The nominal Rabi frequency hovers around \SI{2.7}{\mega \hertz}, yet due to 
the de-tuning induced by the micro-magnet's field, the Rabi frequency escalates to \SI{2.85}{\mega \hertz} at the micro-magnet's maximal field point. This increase in Rabi frequency, corresponding 
to a de-tuning of approximately \SI{0.9}{\mega \hertz}, as can be deduced from Eq. \eqref{eq:general_rabi}.

In Fig.\,\ref{fig:inhomo}(b), we present the corresponding simulated Rabi frequency map for 
a micro-magnet. To comprehend the impact of de-tuning on the sensed magnetic 
field, we compute the phase acquired as a function of phase shift $\delta$ under 
conditions of perfect and imperfect $\pi$ pulses. Fig.\,\ref{fig:inhomo}(c) 
displays the simulated signal for perfect $\pi$ and $\pi/2$ pulses under an 
applied AC magnetic field of \SI{3.5}{\micro\tesla}, and a frequency of \SI{300}
{\kilo\hertz}. As anticipated, fitting to Eq.\,\eqref{eq:xy_signal} yields the 
expected \SI{3.5}{\micro\tesla}. However, with an applied de-tuning of \SI{0.9}
{\mega\hertz}, the acquired net phase diminishes, resulting in an erroneously 
sensed AC signal of \SI{3.25}{\micro\tesla}. This discrepancy arises from the 
imperfect nature of the $\pi$ and $\pi/2$ pulses, which only elevate the 
\ce{NV-} centers to $0.9\ket{1}$ instead of the excited state $\ket{1}$. 
Regardless of whether the signal is de-tuned to the right or left, the net sensed 
signal is consistently less than the applied signal, as depicted in Fig.\,\ref{fig:inhomo}(d) \cite{PhysRevA.99.012110}. This observation aligns with Eq. 
\eqref{eq:general_rabi}. In Fig.\,\ref{fig:inhomo}(e), we illustrate the error in 
the sensed signal due to imperfect pulses. Notably, negligible error is evident 
until approximately \SI{0.5}{\mega\hertz}, beyond which the error escalates 
significantly, reaching nearly $20\%$ for a de-tuning of \SI{2}{\mega\hertz}. 
It's worth noting that the error consistently manifests negatively, indicating 
that the sensed field is consistently underestimated compared to the actual 
applied field. Fig.\,\ref{fig:inhomo}(f) presents the AC simulated magnetic field 
map from a micro-magnet, while Fig.\,\ref{fig:inhomo}(g) illustrates the sensed 
magnetic field map due to $\pi$ pulse errors. The positive lobe of the field 
remains mostly unaffected since most de-tuning in this region is below \SI{0.5}
{\mega\hertz}. However, certain parts of the negative lobe exhibit more negative 
values than actual. Nevertheless, the slope of the line connecting the negative 
lobe and the positive lobe remains nearly constant, serving as the basis for 
estimating the change in dipole moment due to applied AC excitation (refer to 
Appendix\,\ref{sec:micro_model}). Another imperfection may arise from the 
background AC magnetic field originating from the static magnetization of the 
micro-magnet. The off-axis DC field from the micro-magnet can generate an AC 
field proportional to the micro-magnet's magnetization.
The AC magnetic field acquired due to 
such off-axis field is given by\,\cite{PRXQuantum.2.030352}
\begin{equation}
 \tilde{B}_{\perp} = \frac{3\gamma_e}{D_\circ}\left(B_{NV\perp} \times \tilde{B}_{NV\perp} \right),'
 \label{eq:perp}
\end{equation}
where $B_{NV\perp}$ is static perpendicular magnetic field, 
and $\tilde{B}_{NV\perp}$ is the AC perpendicular magnetic 
field.  In our case, $\tilde{B}_{NV\perp}$ is around
\SI{2}{\micro\tesla}, and $B_{NV\perp}$ is around
\SI{40}{\micro\tesla}. Substituting these values into
equation Eq.\,\eqref{eq:perp}, we get a field 
$\tilde{B}_{\perp}$ of around \SI{2}{\nano \tesla}, which is 
far lower than the AC field from the micro-magnet and hence we 
can neglect it.

\begin{figure}
 \centering
 \includegraphics[width=\linewidth]{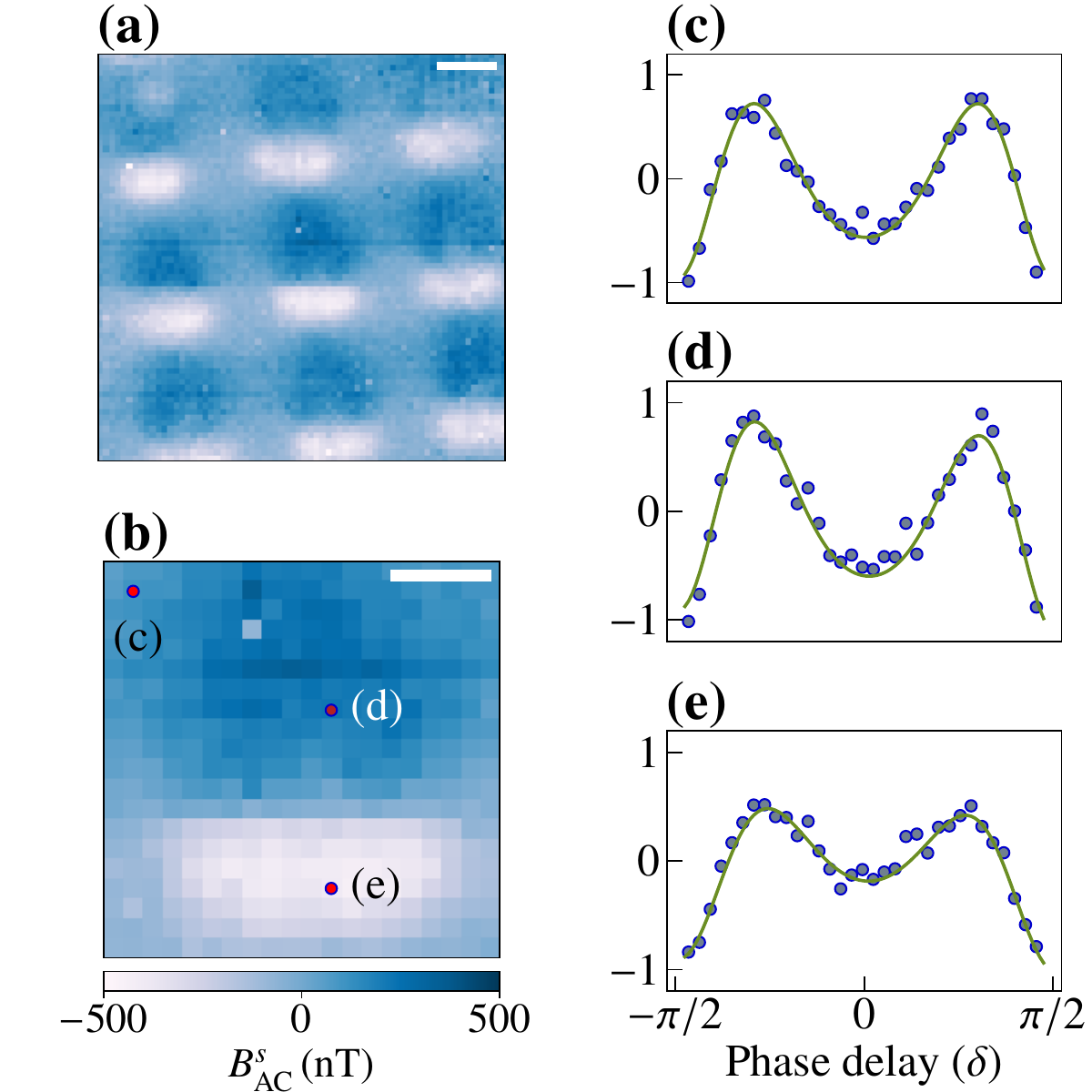}
 \caption{AC magnetic field from micro-magnets. (a). Wide-field image of
 AC magnetic field image from the micro-magnets. The micro-magnets were excited
 with a field of amplitude \SI{3.5}{\micro \tesla} and a frequency of 
 \SI{300}{\kilo \hertz}. The sensing is done using the XY-8 DD sequence. The image is acquired by
 averaging three datasets each acquired for a time duration of \SI{15}{\minute}. The scale-bar is \SI{10}{\um}.  (b). Zoomed in 
 image of the central micro-magnet. The scale-bar is \SI{5}{\um}.  
 (c), (d), (e)  The XY-8 phase delay signal at pixels labeled (c), (d), and (e) in (b) fit to 
 Eq.\,\eqref{eq:xy_signal}.}
 \label{fig:ac_susc}
\end{figure}

Fig.\,\ref{fig:ac_susc}(a) displays the resultant AC magnetic field map for an 
excitation of \SI{3.5}{\micro\tesla} at a frequency of \SI{300}{\kilo\hertz}. In 
our experiments, we employ the XY-4 sequence for excitation frequencies of 
\SI{200}{\kilo\hertz} and \SI{250}{\kilo\hertz}, while adopting the XY-8 sequence 
for higher frequencies. The field map depicted in Fig.\,\ref{fig:ac_susc}(a), and 
subsequent maps have had the effect of the excitation field removed. Fig.\,\ref{fig:ac_susc}(b) zooms in on the AC magnetic field from the central 
micro-magnet. Both positive and negative lobes are clearly discernible, indicating 
that the field map is not solely a result of DC de-tuning but rather represents 
an actual AC magnetic field emanating from the micromagnet. Fig.\,\ref{fig:ac_susc}(c) to Fig.\,\ref{fig:ac_susc}(e) illustrate the contrast variation as a function of 
phase delay $\delta$, fitted to Eq.\,\eqref{eq:xy_signal} for points labeled in 
Fig.\,\ref{fig:ac_susc}(b).

\begin{figure}
\centering
\includegraphics[width=\linewidth]{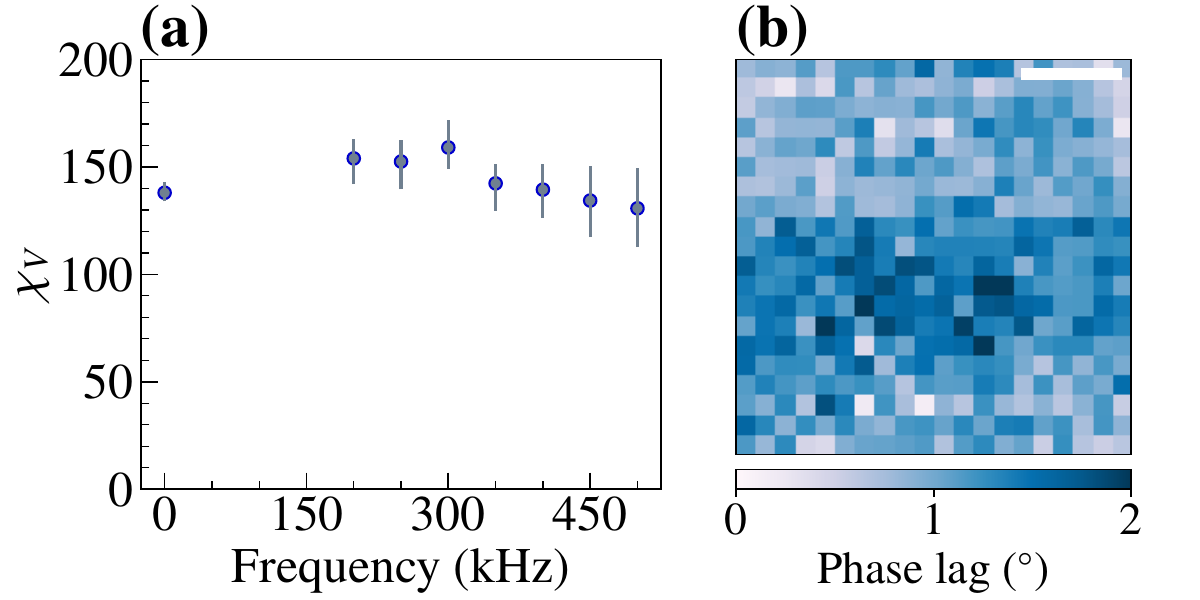}
\label{fig:susceptibility}
\caption{AC susceptibility. (a). Real part of the AC susceptibility at various frequencies. 
(b). Phase relative to the excitation signal around the central micro-magnet at 
\SI{3.5}{\micro\tesla} excitation and \SI{300}{\kilo \hertz} frequency. The scale-bar is \SI{5}{\um}.} 
\label{fig:susceptibility_phase}
\end{figure}

In Fig.\,\ref{fig:susceptibility_phase}(a), the volume-normalized susceptibility $
\chi_V$ is depicted for various applied frequencies of up to \SI{500}{\kilo\hertz}. 
It's noteworthy that $\chi_V$ closely mirrors the DC susceptibility across this 
frequency range due to the rapid movement of domain walls in \ce{Py}, 
facilitating the magnetization's quick alignment with the applied magnetic field 
within a very short period of time compared to the period of the excitation signal. An essential aspect of AC susceptibility measurements is 
considering the phase of the sample's signal with respect to the excitation field. 
When the AC excitation period  is similar to  the relaxation time of the 
magnetic material, the magnetization lags behind the magnetic excitation. 
Assuming the magnetization response adheres to a damped harmonic oscillator 
model, nearing resonance with the AC excitation frequency results in a reduction 
of the in-phase component of magnetization, leaving predominantly the out-of-
phase component. Consequently, the field emitted by the micro-magnet undergoes a 
$90^\circ$ phase shift with respect to  the excitation signal.

Fig.\,\ref{fig:susceptibility_phase}(b) illustrates the phase variation surrounding a micro-magnet 
at an excitation frequency of \SI{300}{\kilo\hertz}. The majority of pixels in the vicinity of the 
micro-magnet display phase fluctuations ranging from $1^\circ$ to $2^\circ$, indicating that noise 
is the primary contributor to this minor fluctuation. This presented phase reflects the overall 
signal phase with respect to the XY-N sequence. Given that the overall signal phase demonstrates no 
notable deviation from the XY-N signal, we infer that the micro-magnet's field similarly exhibits 
negligible phase shifts (see Appendix\,\ref{sec:sensitivity} for additional analysis).

\section{Conclusion and Outlook}
In summary, our study successfully demonstrated the measurement of micro-scale AC susceptibility utilizing \ce{NV-} centers in diamond within a widefield configuration employing the XY dynamical decoupling sequences. Through the excitation of micro-magnets with AC magnetic fields, we were able to detect the stray field emanating from these micro-magnets, corresponding to a change of approximately \SI{1}{\femto\joule\per\tesla} of magnetic moment. Moreover, leveraging our pixel resolution of \SI{1}{\micro\meter} and a field of view spanning \SI{70}{\micro\meter}, we effectively mapped the AC susceptibility across an array of \num{9} micro-magnets spaced at intervals of \SI{25}{\micro\meter}. Additionally, our methodology facilitated the measurement of the phase of the magnetic field originating from the micro-magnets, thanks to the phase-sensitive sensing protocol. Furthermore, we conducted an analysis to identify sources of non-idealities, pinpointing static inhomogeneity as a significant factor. This static inhomogeneity contributes to errors in the $\pi$ pulses within the sensing protocol, underscoring the importance of addressing such challenges for enhanced accuracy in future measurements.

\section*{Acknowledgement}
K.S. acknowledges financial support from DST Quest, SERB Power Research Grant, AOARD grant FA2386-23-1-4012 and I-Hub Divya Sampark Grant in addition to support from IITB Nano-Fabrication facility for sample preparation. 
  
\section*{Conflict of Interest}
The authors have no conflicts to disclose.

\section*{Data Availability}
The data that support the findings of this work is available upon reasonable 
request from the authors.

\appendix
\section{Experimental Setup}\label{sec:exp_details}
We use a home built widefield microscope for the experiments. 
The diamond used is a square of side \SI{4}{\mm}, and thickness
\SI{250}{\um}, and is provided by Element six. The inherent $T_2$ of the diamond 
measured using an XY-8 pulse sequence is \SI{21}{\micro \second}. Hence, we keep the length
of our sequence near this value. For example, the length of an XY-8 sequence for sensing a 
frequency of 
\SI{300}{\kilo \hertz} signal,  including micro-wave pulses, is \SI{14.8}{\micro \second}. 
Moreover, the minimum frame rate of the camera is \SI{2.5}{\kilo \hertz}. This restriction
demands the length of the sequence to be less than \SI{100}{\micro \second}.

The diamond is excited by \SI{532}{\nm} diode laser (Sprout H)
via a 100x, \num{0.9} NA objective (Olympus MPlanFL N), and the resultant
fluorescence is collected through the same objective. The incident
laser power is \SI{300}{\milli \watt}. The red fluorescence collected 
passes through band pass filters (\SI{690}{\nm} to \SI{800}{\nm})
and is recorded by the lockin camera (Heliotis C3). The camera records at a 
frame rate \SI{8}{\kilo \hertz}. The microwave signals are provided by
by a microwave signal generator (SRS386), and amplified by an
an amplifier (Mini-Circuits ZHL-16W-43-S+). DC to  microwave switches
(Mini-Circuits  ZASWA-2-50DR+) are used to switch the micro-waves as well as to provide
switching for the IQ modulation.  The phases of the pulses are controlled 
by the in-built IQ 
modulation of the signal generator, as described 
in\,\cite{Bucher2019}.   The AC excitation is provided by a \SI{4}{\mm} wide copper strip 
which is terminated by two \SI{3}{\ohm}, and \SI{1}{\watt} rated resistors 
using a function generator (Textronix AFG1022). The pulses are provided by
Pulseblaster ESRPro500 pulse generator. 

The permalloy samples were prepared by patterning \SI{7}{\um} diameter 
disks on \ce{LiNbO3} substrate using photo-lithography, and then sputtering a stack 
of \ce{Ti}/\ce{Ni_{0.8}Fe_{0.2}}/\ce{Pt} (\SI{5}{\nm}/\SI{30}{\nm}/\SI{10}{\nm}) on it. However, after deposition, we
found the diameter of the micro-magnets to have shrunken to \SI{5.8\pm 0.4}{\micro \tesla}. 
We used 
\ce{LiNbO3} as the substrate because it is an insulating material. We found that any 
conducting surface near the \ce{NV-} layer degrades the micro-wave power, in-turn,
reducing the Rabi oscillation frequency.

\section{Magnitude and Phase Sensitivity}\label{sec:sensitivity}

\begin{figure}  
\centering
 \includegraphics[width=0.7\linewidth]{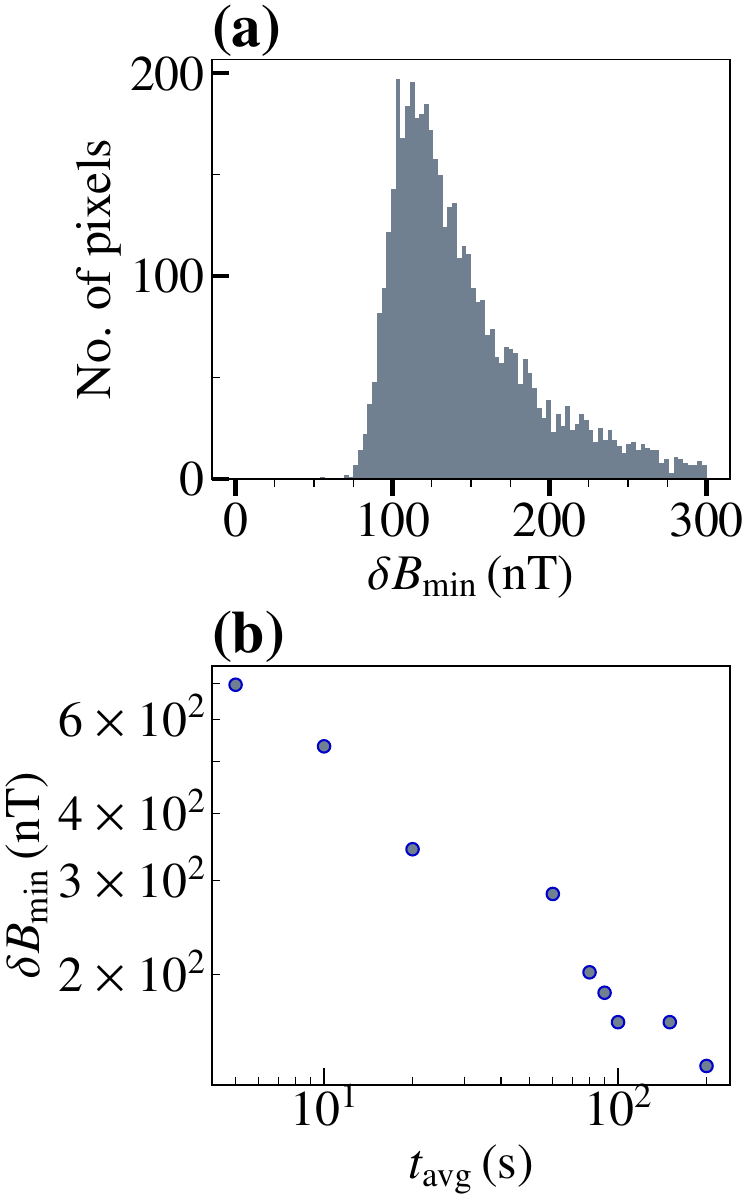} 
 \caption{Magnitude and phase sensitivity. (a). Histogram of the number of minimum detectable 
 magnetic field $\delta B_{\text{min}}$  defined in Eq.\,\eqref{eq:min_magnitude} across the 
 pixels for an averaging time of \SI{200}{\second}. (b). Dependence of median minimum detectable 
 signal with averaging time $t_{\text{avg}}$. }
 \label{fig:sensitivity}    
\end{figure}

\begin{figure*}[ht]   
\centering
 \includegraphics[width=\linewidth]{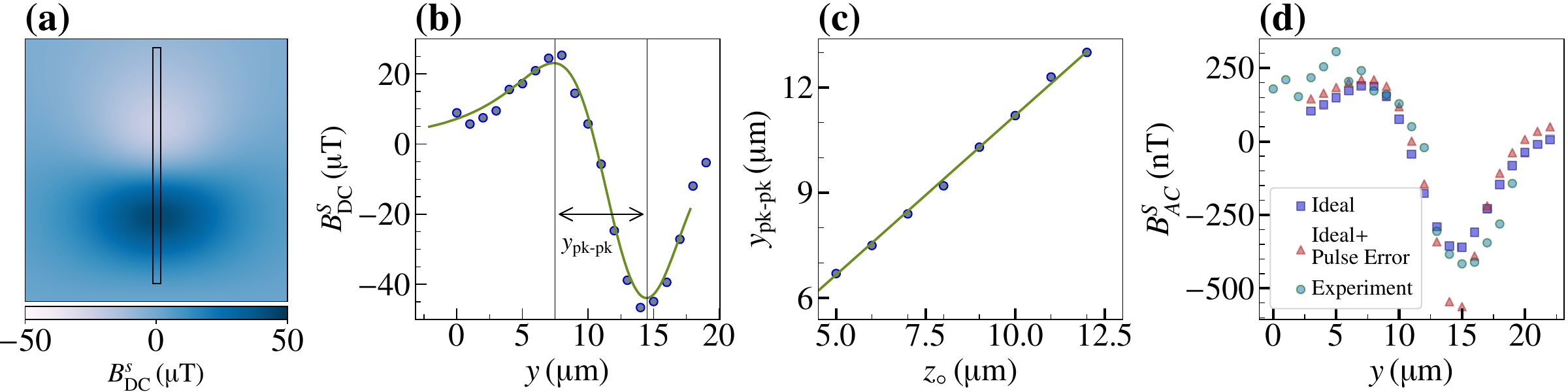}
 \caption{(a). Calculated stray magnetic field for a micro-
 magnet at a stand-off distance of \SI{5.5}{\um}. The rectangular box indicates the line-cut at \SI{1}{\milli \tesla} applied field (See Fig.\,\ref{fig:DC}(b) for experimental map). (b). Line cut 
 along the dipole. The data points are from the experiment magnetic field map in Fig.\,\ref{fig:DC}(b). (c). Peak to peak distance $y_{\text{pk-pk}}$ as a function of stand-off distance $z_\circ$ fit to a straight line. (d). The 
 blue dots represent line cut of the simulated AC field from the sample
 (see Fig.\,\ref{fig:inhomo}(f)). The green dots represent the line cut from 
 Fig.\,\ref{fig:ac_susc}(b), and finally the orange dots represent the line cut
 from Fig.\,\ref{fig:inhomo}(g).}
 \label{fig:magnetic_model}    
\end{figure*} 

The magnitude and phase sensitivity can be derived from Eq.\,\eqref{eq:amp_sensitivity},
and Eq.\,\eqref{eq:phase_sensitivity}. The minimum magnitude of AC field that can be 
sensed is given by 
\begin{equation}
    \Delta B_{\text{AC}}^{\text{min}} =  \frac{\sigma_{S_{XY}}}{\abs{\pdv{S_{XY}}{B_{\text{AC}}}}_{\text{max}}} = \frac{\sigma_{S_{\text{XY}}}}{C\kappa}
    \label{eq:min_magnitude}
\end{equation}
where $\sigma_{S_{XY}}$ is the standard deviation in the XY signal for a fixed initial phase.  
From 
Eq.\,\eqref{eq:amp_sensitivity}, it
can be seen that the maximum slope occurs when the initial phase $\delta = 0$, and 
$\kappa B_{\text{AC}} = (2n+1)\pi/2$. 
In Fig.\,\ref{fig:sensitivity}(a), the histogram of the minimum detectable field  as
defined in Eq.\,\eqref{eq:min_magnitude} for an applied AC magnetic field of 
\SI{5}{\micro \tesla} and an acquisition time of \SI{200}{\s} is shown. The median value of 
$\Delta B_{\text{min}}$ is \SI{135}{\nano \tesla}, with around 80\% of the pixels lying between
\SI{75}{\nano \tesla} to \SI{180}{\nano\tesla}. Moreover, for an averaging time of \SI{5}{\second},
the median minimum detectable field is \SI{630}{\nano \tesla} as shown 
in Fig.\,\ref{fig:sensitivity}(b).

The phase sensitivity, as can be seen from Eq.\,\eqref{eq:phase_sensitivity},  is 
intricately connected to the amplitude of the AC field $B_{AC}$. The minimum phase 
that can be detected around $\delta = (2n+1)\pi/2$ is given by
\begin{equation}
    \Delta \delta_{\text{min}} = \frac{\sigma_{S_{\text{XY}}}}
    {\abs{\pdv{S_{\text{XY}}}{B_{AC}}}_{\text{max}}} = \frac{\sigma_{S_{\text{XY}}}}
    {C\cdot B_{\text{AC}} \cdot \kappa}.
    \label{eq:min_phase}
\end{equation}
The equation for minimum detectable phase as can be seen from Eq.\,\eqref{eq:min_phase} is similar 
to 
the the equation for minimum detectable magnitude in Eq.\,\eqref{eq:min_magnitude}, but with a 
scaling
factor of $B_{AC}$. Thus, for \SI{200}{\second} averaging, and an applied excitation of 
\SI{5}{\micro \second}, the minimum detectable phase of around $1.6^\circ$. 

However, an important point to note here is 
that in the
case of AC susceptibility, we are  mainly interested in measuring the phase of 
the field from 
the micro-magnets with respect to the excitation field \emph{rather than} the 
phase of the total
AC signal with respect to the first $\pi/2$ pulse. Let us consider the case  
where, due the
excitation given in Eq.\,\eqref{eq:Bac}, the field from the sample is 
\begin{equation}
    b_{\text{AC}}^{s}(t) = B_{\text{AC}}^{s} \cos\left( 2\pi f_{ac} t + \delta + \delta_s \right),
\end{equation}
then the total field $B_{\text{AC}}$ will experience an additional 
phase shift $\delta_{\text{add}}$ from $\delta$ that is equal to 
\begin{equation}
    \tan \delta_{\text{add}} =  \frac{B_{\text{AC}}^s \sin  \delta_s}{B_{\text{AC}}^a + B_{\text{AC}}^s\cos \delta_s}.
    \label{eq:net_phase}
\end{equation}
For a case where $B_{\text{AC}}^a \gg  B_{\text{AC}}^s $, and a shift in-phase that is close to zero, we get
a relationship
\begin{equation}
\delta_{\text{add}} = \frac{B_{\text{AC}}^s}{B_{\text{AC}}^a} \delta_\circ.
\label{eq:susceptibility_phase_shift}
\end{equation}
Combining Eq.\,\eqref{eq:susceptibility_phase_shift}, and Eq.\,\eqref{eq:min_phase}, we get
\begin{equation}
        \Delta \delta_{s}^{\text{min}} = \frac{\sigma_{S_{\text{XY}}}}
    {C\cdot B_{\text{AC}}^{s} \cdot \kappa}.
\end{equation}
In our case the maximum field from the micro-magnet is around \num{10} times smaller than the 
excitation field. The minimum detectable phase in this case would be $16^\circ$. However, because,
the AC field from the micro-magnet is not uniform, the minimum detectable phase would also
not be uniform. 
Another case of significant importance is the case where 
$\delta_\circ \to \pi/2$. In such a case, using Taylor expansion around $\pi/2$, 
and ignoring quadratic terms, we get from Eq.\,\eqref{eq:net_phase} the net phase
acquired is 
\begin{equation}
    \delta_{\text{add}} = \frac{B_{AC}^s}{B_{AC}^a}.
\end{equation}
For a perspective if $B_{AC}^s/B_{AC}^a = 1/10$, a phase shift of $90^\circ$ from the sample 
signal will
result in a phase shift of $5.7^\circ$ in the total signal.

\section{Micro-magnet Model}\label{sec:micro_model}

The formulas to calculate the  stray magnetic field from a 
cylindrical magnet  of  arbitrary uniform  magnetization are 
provided in Ref.\,\cite{CACIAGLI2018423}. We use these formulas 
to extract the dipole moment of the micro-magnet from the stray 
magnetic-field. First, we calculate the stray magnetic 
field profile of a \SI{5.8}{\um} diameter micro-magnet of a given 
dipole moment for various stand-off distances. The projection of 
the stray magnetic field along the \ce{NV-} axis is given by
$B_{\text{NV}} = \sqrt{2/3}\,B_y + \sqrt{1/3}\,B_z$\,
\cite{PhysRevApplied.10.034044}. The peak-to-peak distance
on the line-cut along the center of the field map is indicative 
of the stand-off distance, 
and the slope of the line joining the peak-to-peak is directly 
proportional to the magnitude of dipole moment. 
In Fig.\,\ref{fig:magnetic_model}(a), the calculated magnetic
field map for a cylinder along an \ce{NV-} axis for a dipole 
moment of \SI{107}{\femto \joule \per \tesla}, and a stand-off 
distance of \SI{6}{\um} is shown. The line-cut is shown in
Fig.\,\ref{fig:magnetic_model}(b) with the experimental data from Fig.\,\ref{fig:DC}(c).
In Fig.\,\ref{fig:magnetic_model}(c), the peak-to-peak distance
is plotted against the stand-off distance. For a point
sized dipole object, the stand-off distance would be the 
equal to the peak-to-peak distance. However, in the case of an 
extended dipole, the relationship would not be perfectly linear.
We obtain the following linear fit for the relationship between 
the peak-to-peak distance and the stand-off distance
\begin{equation}
 y_{\text{pk-pk}} = 2.12 + 0.9 z_\circ,
 \label{eq:stand_off}
\end{equation}
where $y_{\text{pk-pk}}\,$(\SI{}{\um}) is the peak-to-peak 
distance, 
and $z_\circ\,$(\SI{}{\um}) is the stand-off distance. The slope increases by 
\SI{14.3}{\micro \tesla \per \um} for an increase in the dipole moment of \SI{100}{\femto \joule \per \tesla}, at the previously stated stand-off distance. 
In Fig.\,\ref{fig:magnetic_model}(d), the line cuts for AC fields from the 
micro-magnets is shown for a frequency of \SI{300}{\kilo \hertz}. The 
experimental data fits well with the ideal simulated data, and does not show
a pronounced peak at the negative lobe as predicted by the error in the de-tuning. However, as can be seen, the slope of all the three lines is very 
close to each other.


\begin{thebibliography}{38}%
\makeatletter
\providecommand \@ifxundefined [1]{%
 \@ifx{#1\undefined}
}%
\providecommand \@ifnum [1]{%
 \ifnum #1\expandafter \@firstoftwo
 \else \expandafter \@secondoftwo
 \fi
}%
\providecommand \@ifx [1]{%
 \ifx #1\expandafter \@firstoftwo
 \else \expandafter \@secondoftwo
 \fi
}%
\providecommand \natexlab [1]{#1}%
\providecommand \enquote  [1]{``#1''}%
\providecommand \bibnamefont  [1]{#1}%
\providecommand \bibfnamefont [1]{#1}%
\providecommand \citenamefont [1]{#1}%
\providecommand \href@noop [0]{\@secondoftwo}%
\providecommand \href [0]{\begingroup \@sanitize@url \@href}%
\providecommand \@href[1]{\@@startlink{#1}\@@href}%
\providecommand \@@href[1]{\endgroup#1\@@endlink}%
\providecommand \@sanitize@url [0]{\catcode `\\12\catcode `\$12\catcode `\&12\catcode `\#12\catcode `\^12\catcode `\_12\catcode `\%12\relax}%
\providecommand \@@startlink[1]{}%
\providecommand \@@endlink[0]{}%
\providecommand \url  [0]{\begingroup\@sanitize@url \@url }%
\providecommand \@url [1]{\endgroup\@href {#1}{\urlprefix }}%
\providecommand \urlprefix  [0]{URL }%
\providecommand \Eprint [0]{\href }%
\providecommand \doibase [0]{http://dx.doi.org/}%
\providecommand \selectlanguage [0]{\@gobble}%
\providecommand \bibinfo  [0]{\@secondoftwo}%
\providecommand \bibfield  [0]{\@secondoftwo}%
\providecommand \translation [1]{[#1]}%
\providecommand \BibitemOpen [0]{}%
\providecommand \bibitemStop [0]{}%
\providecommand \bibitemNoStop [0]{.\EOS\space}%
\providecommand \EOS [0]{\spacefactor3000\relax}%
\providecommand \BibitemShut  [1]{\csname bibitem#1\endcsname}%
\let\auto@bib@innerbib\@empty
\bibitem [{\citenamefont {Ramasubramanian}\ \emph {et~al.}(2020)\citenamefont {Ramasubramanian}, \citenamefont {K{\'a}kay}, \citenamefont {Fowley}, \citenamefont {Yildirim}, \citenamefont {Matthes}, \citenamefont {Sorokin}, \citenamefont {Titova}, \citenamefont {Hilliard}, \citenamefont {B{\"o}ttger}, \citenamefont {H{\"u}bner}, \citenamefont {Gemming}, \citenamefont {Schulz}, \citenamefont {Kronast}, \citenamefont {Makarov}, \citenamefont {Fassbender},\ and\ \citenamefont {Deac}}]{Ramasubramanian2020}%
  \BibitemOpen
  \bibfield  {author} {\bibinfo {author} {\bibfnamefont {L.}~\bibnamefont {Ramasubramanian}}, \bibinfo {author} {\bibfnamefont {A.}~\bibnamefont {K{\'a}kay}}, \bibinfo {author} {\bibfnamefont {C.}~\bibnamefont {Fowley}}, \bibinfo {author} {\bibfnamefont {O.}~\bibnamefont {Yildirim}}, \bibinfo {author} {\bibfnamefont {P.}~\bibnamefont {Matthes}}, \bibinfo {author} {\bibfnamefont {S.}~\bibnamefont {Sorokin}}, \bibinfo {author} {\bibfnamefont {A.}~\bibnamefont {Titova}}, \bibinfo {author} {\bibfnamefont {D.}~\bibnamefont {Hilliard}}, \bibinfo {author} {\bibfnamefont {R.}~\bibnamefont {B{\"o}ttger}}, \bibinfo {author} {\bibfnamefont {R.}~\bibnamefont {H{\"u}bner}}, \bibinfo {author} {\bibfnamefont {S.}~\bibnamefont {Gemming}}, \bibinfo {author} {\bibfnamefont {S.~E.}\ \bibnamefont {Schulz}}, \bibinfo {author} {\bibfnamefont {F.}~\bibnamefont {Kronast}}, \bibinfo {author} {\bibfnamefont {D.}~\bibnamefont {Makarov}}, \bibinfo {author} {\bibfnamefont {J.}~\bibnamefont {Fassbender}}, \ and\ \bibinfo {author}
  {\bibfnamefont {A.}~\bibnamefont {Deac}},\ }\href {\doibase 10.1021/acsami.0c08024} {\bibfield  {journal} {\bibinfo  {journal} {ACS Applied Materials {\&} Interfaces}\ }\textbf {\bibinfo {volume} {12}},\ \bibinfo {pages} {27812} (\bibinfo {year} {2020})}\BibitemShut {NoStop}%
\bibitem [{\citenamefont {Shinjo}\ \emph {et~al.}(2000)\citenamefont {Shinjo}, \citenamefont {Okuno}, \citenamefont {Hassdorf}, \citenamefont {Shigeto},\ and\ \citenamefont {Ono}}]{doi:10.1126/science.289.5481.930}%
  \BibitemOpen
  \bibfield  {author} {\bibinfo {author} {\bibfnamefont {T.}~\bibnamefont {Shinjo}}, \bibinfo {author} {\bibfnamefont {T.}~\bibnamefont {Okuno}}, \bibinfo {author} {\bibfnamefont {R.}~\bibnamefont {Hassdorf}}, \bibinfo {author} {\bibfnamefont {K.}~\bibnamefont {Shigeto}}, \ and\ \bibinfo {author} {\bibfnamefont {T.}~\bibnamefont {Ono}},\ }\href {\doibase 10.1126/science.289.5481.930} {\bibfield  {journal} {\bibinfo  {journal} {Science}\ }\textbf {\bibinfo {volume} {289}},\ \bibinfo {pages} {930} (\bibinfo {year} {2000})},\ \Eprint {http://arxiv.org/abs/https://www.science.org/doi/pdf/10.1126/science.289.5481.930} {https://www.science.org/doi/pdf/10.1126/science.289.5481.930} \BibitemShut {NoStop}%
\bibitem [{\citenamefont {Stueber}\ \emph {et~al.}(2021)\citenamefont {Stueber}, \citenamefont {Villanova}, \citenamefont {Aponte}, \citenamefont {Xiao},\ and\ \citenamefont {Colvin}}]{pharmaceutics13070943}%
  \BibitemOpen
  \bibfield  {author} {\bibinfo {author} {\bibfnamefont {D.~D.}\ \bibnamefont {Stueber}}, \bibinfo {author} {\bibfnamefont {J.}~\bibnamefont {Villanova}}, \bibinfo {author} {\bibfnamefont {I.}~\bibnamefont {Aponte}}, \bibinfo {author} {\bibfnamefont {Z.}~\bibnamefont {Xiao}}, \ and\ \bibinfo {author} {\bibfnamefont {V.~L.}\ \bibnamefont {Colvin}},\ }\href {\doibase 10.3390/pharmaceutics13070943} {\bibfield  {journal} {\bibinfo  {journal} {Pharmaceutics}\ }\textbf {\bibinfo {volume} {13}} (\bibinfo {year} {2021}),\ 10.3390/pharmaceutics13070943}\BibitemShut {NoStop}%
\bibitem [{\citenamefont {Hsing}\ \emph {et~al.}(2007)\citenamefont {Hsing}, \citenamefont {Xu},\ and\ \citenamefont {Zhao}}]{https://doi.org/10.1002/elan.200603785}%
  \BibitemOpen
  \bibfield  {author} {\bibinfo {author} {\bibfnamefont {I.-M.}\ \bibnamefont {Hsing}}, \bibinfo {author} {\bibfnamefont {Y.}~\bibnamefont {Xu}}, \ and\ \bibinfo {author} {\bibfnamefont {W.}~\bibnamefont {Zhao}},\ }\href {\doibase https://doi.org/10.1002/elan.200603785} {\bibfield  {journal} {\bibinfo  {journal} {Electroanalysis}\ }\textbf {\bibinfo {volume} {19}},\ \bibinfo {pages} {755} (\bibinfo {year} {2007})}\BibitemShut {NoStop}%
\bibitem [{\citenamefont {Burch}\ \emph {et~al.}(2018)\citenamefont {Burch}, \citenamefont {Mandrus},\ and\ \citenamefont {Park}}]{Burch2018}%
  \BibitemOpen
  \bibfield  {author} {\bibinfo {author} {\bibfnamefont {K.~S.}\ \bibnamefont {Burch}}, \bibinfo {author} {\bibfnamefont {D.}~\bibnamefont {Mandrus}}, \ and\ \bibinfo {author} {\bibfnamefont {J.-G.}\ \bibnamefont {Park}},\ }\href {\doibase 10.1038/s41586-018-0631-z} {\bibfield  {journal} {\bibinfo  {journal} {Nature}\ }\textbf {\bibinfo {volume} {563}},\ \bibinfo {pages} {47} (\bibinfo {year} {2018})}\BibitemShut {NoStop}%
\bibitem [{\citenamefont {Hirohata}\ \emph {et~al.}(2020)\citenamefont {Hirohata}, \citenamefont {Yamada}, \citenamefont {Nakatani}, \citenamefont {Prejbeanu}, \citenamefont {Diény}, \citenamefont {Pirro},\ and\ \citenamefont {Hillebrands}}]{HIROHATA2020166711}%
  \BibitemOpen
  \bibfield  {author} {\bibinfo {author} {\bibfnamefont {A.}~\bibnamefont {Hirohata}}, \bibinfo {author} {\bibfnamefont {K.}~\bibnamefont {Yamada}}, \bibinfo {author} {\bibfnamefont {Y.}~\bibnamefont {Nakatani}}, \bibinfo {author} {\bibfnamefont {I.-L.}\ \bibnamefont {Prejbeanu}}, \bibinfo {author} {\bibfnamefont {B.}~\bibnamefont {Diény}}, \bibinfo {author} {\bibfnamefont {P.}~\bibnamefont {Pirro}}, \ and\ \bibinfo {author} {\bibfnamefont {B.}~\bibnamefont {Hillebrands}},\ }\href {\doibase https://doi.org/10.1016/j.jmmm.2020.166711} {\bibfield  {journal} {\bibinfo  {journal} {Journal of Magnetism and Magnetic Materials}\ }\textbf {\bibinfo {volume} {509}},\ \bibinfo {pages} {166711} (\bibinfo {year} {2020})}\BibitemShut {NoStop}%
\bibitem [{\citenamefont {Topping}\ and\ \citenamefont {Blundell}(2018)}]{Topping_2019}%
  \BibitemOpen
  \bibfield  {author} {\bibinfo {author} {\bibfnamefont {C.~V.}\ \bibnamefont {Topping}}\ and\ \bibinfo {author} {\bibfnamefont {S.~J.}\ \bibnamefont {Blundell}},\ }\href {\doibase 10.1088/1361-648X/aaed96} {\bibfield  {journal} {\bibinfo  {journal} {Journal of Physics: Condensed Matter}\ }\textbf {\bibinfo {volume} {31}},\ \bibinfo {pages} {013001} (\bibinfo {year} {2018})}\BibitemShut {NoStop}%
\bibitem [{\citenamefont {Cullity}\ and\ \citenamefont {Graham}(2011)}]{cullity2011introduction}%
  \BibitemOpen
  \bibfield  {author} {\bibinfo {author} {\bibfnamefont {B.~D.}\ \bibnamefont {Cullity}}\ and\ \bibinfo {author} {\bibfnamefont {C.~D.}\ \bibnamefont {Graham}},\ }\href@noop {} {\emph {\bibinfo {title} {Introduction to magnetic materials}}}\ (\bibinfo  {publisher} {John Wiley \& Sons},\ \bibinfo {year} {2011})\BibitemShut {NoStop}%
\bibitem [{\citenamefont {Postulka}\ \emph {et~al.}(2019)\citenamefont {Postulka}, \citenamefont {Eibisch}, \citenamefont {Holzmann}, \citenamefont {Wolf},\ and\ \citenamefont {Lang}}]{10.1063/1.5046475}%
  \BibitemOpen
  \bibfield  {author} {\bibinfo {author} {\bibfnamefont {L.}~\bibnamefont {Postulka}}, \bibinfo {author} {\bibfnamefont {P.}~\bibnamefont {Eibisch}}, \bibinfo {author} {\bibfnamefont {A.}~\bibnamefont {Holzmann}}, \bibinfo {author} {\bibfnamefont {B.}~\bibnamefont {Wolf}}, \ and\ \bibinfo {author} {\bibfnamefont {M.}~\bibnamefont {Lang}},\ }\href {\doibase 10.1063/1.5046475} {\bibfield  {journal} {\bibinfo  {journal} {Review of Scientific Instruments}\ }\textbf {\bibinfo {volume} {90}},\ \bibinfo {pages} {033901} (\bibinfo {year} {2019})}\BibitemShut {NoStop}%
\bibitem [{\citenamefont {Toraille}\ \emph {et~al.}(2018)\citenamefont {Toraille}, \citenamefont {A{\"i}zel}, \citenamefont {Balloul}, \citenamefont {Vicario}, \citenamefont {Monzel}, \citenamefont {Coppey}, \citenamefont {Secret}, \citenamefont {Siaugue}, \citenamefont {Sampaio}, \citenamefont {Rohart}, \citenamefont {Vernier}, \citenamefont {Bonnemay}, \citenamefont {Debuisschert}, \citenamefont {Rondin}, \citenamefont {Roch},\ and\ \citenamefont {Dahan}}]{Toraille2018}%
  \BibitemOpen
  \bibfield  {author} {\bibinfo {author} {\bibfnamefont {L.}~\bibnamefont {Toraille}}, \bibinfo {author} {\bibfnamefont {K.}~\bibnamefont {A{\"i}zel}}, \bibinfo {author} {\bibfnamefont {{\'E}.}~\bibnamefont {Balloul}}, \bibinfo {author} {\bibfnamefont {C.}~\bibnamefont {Vicario}}, \bibinfo {author} {\bibfnamefont {C.}~\bibnamefont {Monzel}}, \bibinfo {author} {\bibfnamefont {M.}~\bibnamefont {Coppey}}, \bibinfo {author} {\bibfnamefont {E.}~\bibnamefont {Secret}}, \bibinfo {author} {\bibfnamefont {J.-M.}\ \bibnamefont {Siaugue}}, \bibinfo {author} {\bibfnamefont {J.}~\bibnamefont {Sampaio}}, \bibinfo {author} {\bibfnamefont {S.}~\bibnamefont {Rohart}}, \bibinfo {author} {\bibfnamefont {N.}~\bibnamefont {Vernier}}, \bibinfo {author} {\bibfnamefont {L.}~\bibnamefont {Bonnemay}}, \bibinfo {author} {\bibfnamefont {T.}~\bibnamefont {Debuisschert}}, \bibinfo {author} {\bibfnamefont {L.}~\bibnamefont {Rondin}}, \bibinfo {author} {\bibfnamefont {J.-F.}\ \bibnamefont {Roch}}, \ and\ \bibinfo {author} {\bibfnamefont
  {M.}~\bibnamefont {Dahan}},\ }\href {\doibase 10.1021/acs.nanolett.8b03222} {\bibfield  {journal} {\bibinfo  {journal} {Nano Letters}\ }\textbf {\bibinfo {volume} {18}},\ \bibinfo {pages} {7635} (\bibinfo {year} {2018})}\BibitemShut {NoStop}%
\bibitem [{\citenamefont {Maertz}\ \emph {et~al.}(2010)\citenamefont {Maertz}, \citenamefont {Wijnheijmer}, \citenamefont {Fuchs}, \citenamefont {Nowakowski},\ and\ \citenamefont {Awschalom}}]{10.1063/1.3337096}%
  \BibitemOpen
  \bibfield  {author} {\bibinfo {author} {\bibfnamefont {B.~J.}\ \bibnamefont {Maertz}}, \bibinfo {author} {\bibfnamefont {A.~P.}\ \bibnamefont {Wijnheijmer}}, \bibinfo {author} {\bibfnamefont {G.~D.}\ \bibnamefont {Fuchs}}, \bibinfo {author} {\bibfnamefont {M.~E.}\ \bibnamefont {Nowakowski}}, \ and\ \bibinfo {author} {\bibfnamefont {D.~D.}\ \bibnamefont {Awschalom}},\ }\href {\doibase 10.1063/1.3337096} {\bibfield  {journal} {\bibinfo  {journal} {Applied Physics Letters}\ }\textbf {\bibinfo {volume} {96}},\ \bibinfo {pages} {092504} (\bibinfo {year} {2010})},\ \Eprint {http://arxiv.org/abs/https://pubs.aip.org/aip/apl/article-pdf/doi/10.1063/1.3337096/14426413/092504\_1\_online.pdf} {https://pubs.aip.org/aip/apl/article-pdf/doi/10.1063/1.3337096/14426413/092504\_1\_online.pdf} \BibitemShut {NoStop}%
\bibitem [{\citenamefont {Hsieh}\ \emph {et~al.}(2019)\citenamefont {Hsieh}, \citenamefont {Bhattacharyya}, \citenamefont {Zu}, \citenamefont {Mittiga}, \citenamefont {Smart}, \citenamefont {Machado}, \citenamefont {Kobrin}, \citenamefont {Höhn}, \citenamefont {Rui}, \citenamefont {Kamrani}, \citenamefont {Chatterjee}, \citenamefont {Choi}, \citenamefont {Zaletel}, \citenamefont {Struzhkin}, \citenamefont {Moore}, \citenamefont {Levitas}, \citenamefont {Jeanloz},\ and\ \citenamefont {Yao}}]{doi:10.1126/science.aaw4352}%
  \BibitemOpen
  \bibfield  {author} {\bibinfo {author} {\bibfnamefont {S.}~\bibnamefont {Hsieh}}, \bibinfo {author} {\bibfnamefont {P.}~\bibnamefont {Bhattacharyya}}, \bibinfo {author} {\bibfnamefont {C.}~\bibnamefont {Zu}}, \bibinfo {author} {\bibfnamefont {T.}~\bibnamefont {Mittiga}}, \bibinfo {author} {\bibfnamefont {T.~J.}\ \bibnamefont {Smart}}, \bibinfo {author} {\bibfnamefont {F.}~\bibnamefont {Machado}}, \bibinfo {author} {\bibfnamefont {B.}~\bibnamefont {Kobrin}}, \bibinfo {author} {\bibfnamefont {T.~O.}\ \bibnamefont {Höhn}}, \bibinfo {author} {\bibfnamefont {N.~Z.}\ \bibnamefont {Rui}}, \bibinfo {author} {\bibfnamefont {M.}~\bibnamefont {Kamrani}}, \bibinfo {author} {\bibfnamefont {S.}~\bibnamefont {Chatterjee}}, \bibinfo {author} {\bibfnamefont {S.}~\bibnamefont {Choi}}, \bibinfo {author} {\bibfnamefont {M.}~\bibnamefont {Zaletel}}, \bibinfo {author} {\bibfnamefont {V.~V.}\ \bibnamefont {Struzhkin}}, \bibinfo {author} {\bibfnamefont {J.~E.}\ \bibnamefont {Moore}}, \bibinfo {author} {\bibfnamefont {V.~I.}\
  \bibnamefont {Levitas}}, \bibinfo {author} {\bibfnamefont {R.}~\bibnamefont {Jeanloz}}, \ and\ \bibinfo {author} {\bibfnamefont {N.~Y.}\ \bibnamefont {Yao}},\ }\href {\doibase 10.1126/science.aaw4352} {\bibfield  {journal} {\bibinfo  {journal} {Science}\ }\textbf {\bibinfo {volume} {366}},\ \bibinfo {pages} {1349} (\bibinfo {year} {2019})},\ \Eprint {http://arxiv.org/abs/https://www.science.org/doi/pdf/10.1126/science.aaw4352} {https://www.science.org/doi/pdf/10.1126/science.aaw4352} \BibitemShut {NoStop}%
\bibitem [{\citenamefont {Dasika}\ \emph {et~al.}(2023)\citenamefont {Dasika}, \citenamefont {Parashar},\ and\ \citenamefont {Saha}}]{10.1063/5.0138301}%
  \BibitemOpen
  \bibfield  {author} {\bibinfo {author} {\bibfnamefont {S.}~\bibnamefont {Dasika}}, \bibinfo {author} {\bibfnamefont {M.}~\bibnamefont {Parashar}}, \ and\ \bibinfo {author} {\bibfnamefont {K.}~\bibnamefont {Saha}},\ }\href {\doibase 10.1063/5.0138301} {\bibfield  {journal} {\bibinfo  {journal} {Review of Scientific Instruments}\ }\textbf {\bibinfo {volume} {94}},\ \bibinfo {pages} {053702} (\bibinfo {year} {2023})},\ \Eprint {http://arxiv.org/abs/https://pubs.aip.org/aip/rsi/article-pdf/doi/10.1063/5.0138301/17148139/053702\_1\_5.0138301.pdf} {https://pubs.aip.org/aip/rsi/article-pdf/doi/10.1063/5.0138301/17148139/053702\_1\_5.0138301.pdf} \BibitemShut {NoStop}%
\bibitem [{\citenamefont {Tetienne}\ \emph {et~al.}(2013)\citenamefont {Tetienne}, \citenamefont {Hingant}, \citenamefont {Rondin}, \citenamefont {Rohart}, \citenamefont {Thiaville}, \citenamefont {Roch},\ and\ \citenamefont {Jacques}}]{PhysRevB.88.214408}%
  \BibitemOpen
  \bibfield  {author} {\bibinfo {author} {\bibfnamefont {J.-P.}\ \bibnamefont {Tetienne}}, \bibinfo {author} {\bibfnamefont {T.}~\bibnamefont {Hingant}}, \bibinfo {author} {\bibfnamefont {L.}~\bibnamefont {Rondin}}, \bibinfo {author} {\bibfnamefont {S.}~\bibnamefont {Rohart}}, \bibinfo {author} {\bibfnamefont {A.}~\bibnamefont {Thiaville}}, \bibinfo {author} {\bibfnamefont {J.-F.}\ \bibnamefont {Roch}}, \ and\ \bibinfo {author} {\bibfnamefont {V.}~\bibnamefont {Jacques}},\ }\href {\doibase 10.1103/PhysRevB.88.214408} {\bibfield  {journal} {\bibinfo  {journal} {Phys. Rev. B}\ }\textbf {\bibinfo {volume} {88}},\ \bibinfo {pages} {214408} (\bibinfo {year} {2013})}\BibitemShut {NoStop}%
\bibitem [{\citenamefont {Huang}\ \emph {et~al.}(2023)\citenamefont {Huang}, \citenamefont {Green}, \citenamefont {Zhou}, \citenamefont {Williams}, \citenamefont {Li}, \citenamefont {Lu}, \citenamefont {Djugba}, \citenamefont {Wang}, \citenamefont {Flebus}, \citenamefont {Ni},\ and\ \citenamefont {Du}}]{Huang2023}%
  \BibitemOpen
  \bibfield  {author} {\bibinfo {author} {\bibfnamefont {M.}~\bibnamefont {Huang}}, \bibinfo {author} {\bibfnamefont {J.~C.}\ \bibnamefont {Green}}, \bibinfo {author} {\bibfnamefont {J.}~\bibnamefont {Zhou}}, \bibinfo {author} {\bibfnamefont {V.}~\bibnamefont {Williams}}, \bibinfo {author} {\bibfnamefont {S.}~\bibnamefont {Li}}, \bibinfo {author} {\bibfnamefont {H.}~\bibnamefont {Lu}}, \bibinfo {author} {\bibfnamefont {D.}~\bibnamefont {Djugba}}, \bibinfo {author} {\bibfnamefont {H.}~\bibnamefont {Wang}}, \bibinfo {author} {\bibfnamefont {B.}~\bibnamefont {Flebus}}, \bibinfo {author} {\bibfnamefont {N.}~\bibnamefont {Ni}}, \ and\ \bibinfo {author} {\bibfnamefont {C.~R.}\ \bibnamefont {Du}},\ }\href {\doibase 10.1021/acs.nanolett.3c02129} {\bibfield  {journal} {\bibinfo  {journal} {Nano Letters}\ }\textbf {\bibinfo {volume} {23}},\ \bibinfo {pages} {8099} (\bibinfo {year} {2023})}\BibitemShut {NoStop}%
\bibitem [{\citenamefont {Broadway}\ \emph {et~al.}(2020)\citenamefont {Broadway}, \citenamefont {Scholten}, \citenamefont {Tan}, \citenamefont {Dontschuk}, \citenamefont {Lillie}, \citenamefont {Johnson}, \citenamefont {Zheng}, \citenamefont {Wang}, \citenamefont {Oganov}, \citenamefont {Tian}, \citenamefont {Li}, \citenamefont {Lei}, \citenamefont {Wang}, \citenamefont {Hollenberg},\ and\ \citenamefont {Tetienne}}]{https://doi.org/10.1002/adma.202003314}%
  \BibitemOpen
  \bibfield  {author} {\bibinfo {author} {\bibfnamefont {D.~A.}\ \bibnamefont {Broadway}}, \bibinfo {author} {\bibfnamefont {S.~C.}\ \bibnamefont {Scholten}}, \bibinfo {author} {\bibfnamefont {C.}~\bibnamefont {Tan}}, \bibinfo {author} {\bibfnamefont {N.}~\bibnamefont {Dontschuk}}, \bibinfo {author} {\bibfnamefont {S.~E.}\ \bibnamefont {Lillie}}, \bibinfo {author} {\bibfnamefont {B.~C.}\ \bibnamefont {Johnson}}, \bibinfo {author} {\bibfnamefont {G.}~\bibnamefont {Zheng}}, \bibinfo {author} {\bibfnamefont {Z.}~\bibnamefont {Wang}}, \bibinfo {author} {\bibfnamefont {A.~R.}\ \bibnamefont {Oganov}}, \bibinfo {author} {\bibfnamefont {S.}~\bibnamefont {Tian}}, \bibinfo {author} {\bibfnamefont {C.}~\bibnamefont {Li}}, \bibinfo {author} {\bibfnamefont {H.}~\bibnamefont {Lei}}, \bibinfo {author} {\bibfnamefont {L.}~\bibnamefont {Wang}}, \bibinfo {author} {\bibfnamefont {L.~C.~L.}\ \bibnamefont {Hollenberg}}, \ and\ \bibinfo {author} {\bibfnamefont {J.-P.}\ \bibnamefont {Tetienne}},\ }\href {\doibase
  https://doi.org/10.1002/adma.202003314} {\bibfield  {journal} {\bibinfo  {journal} {Advanced Materials}\ }\textbf {\bibinfo {volume} {32}},\ \bibinfo {pages} {2003314} (\bibinfo {year} {2020})},\ \Eprint {http://arxiv.org/abs/https://onlinelibrary.wiley.com/doi/pdf/10.1002/adma.202003314} {https://onlinelibrary.wiley.com/doi/pdf/10.1002/adma.202003314} \BibitemShut {NoStop}%
\bibitem [{\citenamefont {Meneses}\ \emph {et~al.}(2024)\citenamefont {Meneses}, \citenamefont {Qi}, \citenamefont {Healey}, \citenamefont {You}, \citenamefont {Robertson}, \citenamefont {Scholten}, \citenamefont {Keerthi}, \citenamefont {Harrison}, \citenamefont {Bera}, \citenamefont {Jyothilal}, \citenamefont {Hollenberg}, \citenamefont {Radha},\ and\ \citenamefont {Tetienne}}]{PhysRevB.109.064416}%
  \BibitemOpen
  \bibfield  {author} {\bibinfo {author} {\bibfnamefont {F.}~\bibnamefont {Meneses}}, \bibinfo {author} {\bibfnamefont {R.}~\bibnamefont {Qi}}, \bibinfo {author} {\bibfnamefont {A.~J.}\ \bibnamefont {Healey}}, \bibinfo {author} {\bibfnamefont {Y.}~\bibnamefont {You}}, \bibinfo {author} {\bibfnamefont {I.~O.}\ \bibnamefont {Robertson}}, \bibinfo {author} {\bibfnamefont {S.~C.}\ \bibnamefont {Scholten}}, \bibinfo {author} {\bibfnamefont {A.}~\bibnamefont {Keerthi}}, \bibinfo {author} {\bibfnamefont {G.}~\bibnamefont {Harrison}}, \bibinfo {author} {\bibfnamefont {A.}~\bibnamefont {Bera}}, \bibinfo {author} {\bibfnamefont {H.}~\bibnamefont {Jyothilal}}, \bibinfo {author} {\bibfnamefont {L.~C.~L.}\ \bibnamefont {Hollenberg}}, \bibinfo {author} {\bibfnamefont {B.}~\bibnamefont {Radha}}, \ and\ \bibinfo {author} {\bibfnamefont {J.-P.}\ \bibnamefont {Tetienne}},\ }\href {\doibase 10.1103/PhysRevB.109.064416} {\bibfield  {journal} {\bibinfo  {journal} {Phys. Rev. B}\ }\textbf {\bibinfo {volume} {109}},\ \bibinfo
  {pages} {064416} (\bibinfo {year} {2024})}\BibitemShut {NoStop}%
\bibitem [{\citenamefont {Healey}\ \emph {et~al.}(2022)\citenamefont {Healey}, \citenamefont {Rahman}, \citenamefont {Scholten}, \citenamefont {Robertson}, \citenamefont {Abrahams}, \citenamefont {Dontschuk}, \citenamefont {Liu}, \citenamefont {Hollenberg}, \citenamefont {Lu},\ and\ \citenamefont {Tetienne}}]{Healey2022}%
  \BibitemOpen
  \bibfield  {author} {\bibinfo {author} {\bibfnamefont {A.~J.}\ \bibnamefont {Healey}}, \bibinfo {author} {\bibfnamefont {S.}~\bibnamefont {Rahman}}, \bibinfo {author} {\bibfnamefont {S.~C.}\ \bibnamefont {Scholten}}, \bibinfo {author} {\bibfnamefont {I.~O.}\ \bibnamefont {Robertson}}, \bibinfo {author} {\bibfnamefont {G.~J.}\ \bibnamefont {Abrahams}}, \bibinfo {author} {\bibfnamefont {N.}~\bibnamefont {Dontschuk}}, \bibinfo {author} {\bibfnamefont {B.}~\bibnamefont {Liu}}, \bibinfo {author} {\bibfnamefont {L.~C.~L.}\ \bibnamefont {Hollenberg}}, \bibinfo {author} {\bibfnamefont {Y.}~\bibnamefont {Lu}}, \ and\ \bibinfo {author} {\bibfnamefont {J.-P.}\ \bibnamefont {Tetienne}},\ }\href {\doibase 10.1021/acsnano.2c04132} {\bibfield  {journal} {\bibinfo  {journal} {ACS Nano}\ }\textbf {\bibinfo {volume} {16}},\ \bibinfo {pages} {12580} (\bibinfo {year} {2022})}\BibitemShut {NoStop}%
\bibitem [{\citenamefont {Laraoui}\ and\ \citenamefont {Ambal}(2022)}]{10.1063/5.0091931}%
  \BibitemOpen
  \bibfield  {author} {\bibinfo {author} {\bibfnamefont {A.}~\bibnamefont {Laraoui}}\ and\ \bibinfo {author} {\bibfnamefont {K.}~\bibnamefont {Ambal}},\ }\href {\doibase 10.1063/5.0091931} {\bibfield  {journal} {\bibinfo  {journal} {Applied Physics Letters}\ }\textbf {\bibinfo {volume} {121}},\ \bibinfo {pages} {060502} (\bibinfo {year} {2022})},\ \Eprint {http://arxiv.org/abs/https://pubs.aip.org/aip/apl/article-pdf/doi/10.1063/5.0091931/16480572/060502\_1\_online.pdf} {https://pubs.aip.org/aip/apl/article-pdf/doi/10.1063/5.0091931/16480572/060502\_1\_online.pdf} \BibitemShut {NoStop}%
\bibitem [{\citenamefont {Zhang}\ \emph {et~al.}(2021)\citenamefont {Zhang}, \citenamefont {Wang}, \citenamefont {Tartaglia}, \citenamefont {Ding}, \citenamefont {Gray}, \citenamefont {Burch}, \citenamefont {Tafti},\ and\ \citenamefont {Zhou}}]{PRXQuantum.2.030352}%
  \BibitemOpen
  \bibfield  {author} {\bibinfo {author} {\bibfnamefont {X.-Y.}\ \bibnamefont {Zhang}}, \bibinfo {author} {\bibfnamefont {Y.-X.}\ \bibnamefont {Wang}}, \bibinfo {author} {\bibfnamefont {T.~A.}\ \bibnamefont {Tartaglia}}, \bibinfo {author} {\bibfnamefont {T.}~\bibnamefont {Ding}}, \bibinfo {author} {\bibfnamefont {M.~J.}\ \bibnamefont {Gray}}, \bibinfo {author} {\bibfnamefont {K.~S.}\ \bibnamefont {Burch}}, \bibinfo {author} {\bibfnamefont {F.}~\bibnamefont {Tafti}}, \ and\ \bibinfo {author} {\bibfnamefont {B.~B.}\ \bibnamefont {Zhou}},\ }\href {\doibase 10.1103/PRXQuantum.2.030352} {\bibfield  {journal} {\bibinfo  {journal} {PRX Quantum}\ }\textbf {\bibinfo {volume} {2}},\ \bibinfo {pages} {030352} (\bibinfo {year} {2021})}\BibitemShut {NoStop}%
\bibitem [{\citenamefont {Robertson}\ \emph {et~al.}(2022)\citenamefont {Robertson}, \citenamefont {Tan}, \citenamefont {Scholten}, \citenamefont {Healey}, \citenamefont {Abrahams}, \citenamefont {Zheng}, \citenamefont {Manchon}, \citenamefont {Wang},\ and\ \citenamefont {Tetienne}}]{Robertson_2023}%
  \BibitemOpen
  \bibfield  {author} {\bibinfo {author} {\bibfnamefont {I.~O.}\ \bibnamefont {Robertson}}, \bibinfo {author} {\bibfnamefont {C.}~\bibnamefont {Tan}}, \bibinfo {author} {\bibfnamefont {S.~C.}\ \bibnamefont {Scholten}}, \bibinfo {author} {\bibfnamefont {A.~J.}\ \bibnamefont {Healey}}, \bibinfo {author} {\bibfnamefont {G.~J.}\ \bibnamefont {Abrahams}}, \bibinfo {author} {\bibfnamefont {G.}~\bibnamefont {Zheng}}, \bibinfo {author} {\bibfnamefont {A.}~\bibnamefont {Manchon}}, \bibinfo {author} {\bibfnamefont {L.}~\bibnamefont {Wang}}, \ and\ \bibinfo {author} {\bibfnamefont {J.-P.}\ \bibnamefont {Tetienne}},\ }\href {\doibase 10.1088/2053-1583/acab73} {\bibfield  {journal} {\bibinfo  {journal} {2D Materials}\ }\textbf {\bibinfo {volume} {10}},\ \bibinfo {pages} {015023} (\bibinfo {year} {2022})}\BibitemShut {NoStop}%
\bibitem [{\citenamefont {de~Gille}\ \emph {et~al.}(2021)\citenamefont {de~Gille}, \citenamefont {McCoey}, \citenamefont {Hall}, \citenamefont {Tetienne}, \citenamefont {Malkemper}, \citenamefont {Keays}, \citenamefont {Hollenberg},\ and\ \citenamefont {Simpson}}]{doi:10.1073/pnas.2112749118}%
  \BibitemOpen
  \bibfield  {author} {\bibinfo {author} {\bibfnamefont {R.~W.}\ \bibnamefont {de~Gille}}, \bibinfo {author} {\bibfnamefont {J.~M.}\ \bibnamefont {McCoey}}, \bibinfo {author} {\bibfnamefont {L.~T.}\ \bibnamefont {Hall}}, \bibinfo {author} {\bibfnamefont {J.-P.}\ \bibnamefont {Tetienne}}, \bibinfo {author} {\bibfnamefont {E.~P.}\ \bibnamefont {Malkemper}}, \bibinfo {author} {\bibfnamefont {D.~A.}\ \bibnamefont {Keays}}, \bibinfo {author} {\bibfnamefont {L.~C.~L.}\ \bibnamefont {Hollenberg}}, \ and\ \bibinfo {author} {\bibfnamefont {D.~A.}\ \bibnamefont {Simpson}},\ }\href {\doibase 10.1073/pnas.2112749118} {\bibfield  {journal} {\bibinfo  {journal} {Proceedings of the National Academy of Sciences}\ }\textbf {\bibinfo {volume} {118}},\ \bibinfo {pages} {e2112749118} (\bibinfo {year} {2021})},\ \Eprint {http://arxiv.org/abs/https://www.pnas.org/doi/pdf/10.1073/pnas.2112749118} {https://www.pnas.org/doi/pdf/10.1073/pnas.2112749118} \BibitemShut {NoStop}%
\bibitem [{\citenamefont {Xu}\ \emph {et~al.}(2024)\citenamefont {Xu}, \citenamefont {Zhang}, \citenamefont {Ma}, \citenamefont {Hou}, \citenamefont {Li}, \citenamefont {Denisenko}, \citenamefont {Li}, \citenamefont {Spatz}, \citenamefont {Wrachtrup}, \citenamefont {Lei}, \citenamefont {Cao}, \citenamefont {Wei},\ and\ \citenamefont {Chu}}]{doi:10.1126/sciadv.adi5300}%
  \BibitemOpen
  \bibfield  {author} {\bibinfo {author} {\bibfnamefont {F.}~\bibnamefont {Xu}}, \bibinfo {author} {\bibfnamefont {S.}~\bibnamefont {Zhang}}, \bibinfo {author} {\bibfnamefont {L.}~\bibnamefont {Ma}}, \bibinfo {author} {\bibfnamefont {Y.}~\bibnamefont {Hou}}, \bibinfo {author} {\bibfnamefont {J.}~\bibnamefont {Li}}, \bibinfo {author} {\bibfnamefont {A.}~\bibnamefont {Denisenko}}, \bibinfo {author} {\bibfnamefont {Z.}~\bibnamefont {Li}}, \bibinfo {author} {\bibfnamefont {J.}~\bibnamefont {Spatz}}, \bibinfo {author} {\bibfnamefont {J.}~\bibnamefont {Wrachtrup}}, \bibinfo {author} {\bibfnamefont {H.}~\bibnamefont {Lei}}, \bibinfo {author} {\bibfnamefont {Y.}~\bibnamefont {Cao}}, \bibinfo {author} {\bibfnamefont {Q.}~\bibnamefont {Wei}}, \ and\ \bibinfo {author} {\bibfnamefont {Z.}~\bibnamefont {Chu}},\ }\href {\doibase 10.1126/sciadv.adi5300} {\bibfield  {journal} {\bibinfo  {journal} {Science Advances}\ }\textbf {\bibinfo {volume} {10}},\ \bibinfo {pages} {eadi5300} (\bibinfo {year} {2024})},\ \Eprint
  {http://arxiv.org/abs/https://www.science.org/doi/pdf/10.1126/sciadv.adi5300} {https://www.science.org/doi/pdf/10.1126/sciadv.adi5300} \BibitemShut {NoStop}%
\bibitem [{\citenamefont {Kayci}\ \emph {et~al.}(2021)\citenamefont {Kayci}, \citenamefont {Fan}, \citenamefont {Bakirman},\ and\ \citenamefont {Herrmann}}]{doi:10.1073/pnas.2112664118}%
  \BibitemOpen
  \bibfield  {author} {\bibinfo {author} {\bibfnamefont {M.}~\bibnamefont {Kayci}}, \bibinfo {author} {\bibfnamefont {J.}~\bibnamefont {Fan}}, \bibinfo {author} {\bibfnamefont {O.}~\bibnamefont {Bakirman}}, \ and\ \bibinfo {author} {\bibfnamefont {A.}~\bibnamefont {Herrmann}},\ }\href {\doibase 10.1073/pnas.2112664118} {\bibfield  {journal} {\bibinfo  {journal} {Proceedings of the National Academy of Sciences}\ }\textbf {\bibinfo {volume} {118}},\ \bibinfo {pages} {e2112664118} (\bibinfo {year} {2021})},\ \Eprint {http://arxiv.org/abs/https://www.pnas.org/doi/pdf/10.1073/pnas.2112664118} {https://www.pnas.org/doi/pdf/10.1073/pnas.2112664118} \BibitemShut {NoStop}%
\bibitem [{\citenamefont {Parashar}\ \emph {et~al.}(2022)\citenamefont {Parashar}, \citenamefont {Bathla}, \citenamefont {Shishir}, \citenamefont {Gokhale}, \citenamefont {Bandyopadhyay},\ and\ \citenamefont {Saha}}]{Parashar2022}%
  \BibitemOpen
  \bibfield  {author} {\bibinfo {author} {\bibfnamefont {M.}~\bibnamefont {Parashar}}, \bibinfo {author} {\bibfnamefont {A.}~\bibnamefont {Bathla}}, \bibinfo {author} {\bibfnamefont {D.}~\bibnamefont {Shishir}}, \bibinfo {author} {\bibfnamefont {A.}~\bibnamefont {Gokhale}}, \bibinfo {author} {\bibfnamefont {S.}~\bibnamefont {Bandyopadhyay}}, \ and\ \bibinfo {author} {\bibfnamefont {K.}~\bibnamefont {Saha}},\ }\href {\doibase 10.1038/s41598-022-12609-3} {\bibfield  {journal} {\bibinfo  {journal} {Scientific Reports}\ }\textbf {\bibinfo {volume} {12}},\ \bibinfo {pages} {8743} (\bibinfo {year} {2022})}\BibitemShut {NoStop}%
\bibitem [{\citenamefont {Levine}\ \emph {et~al.}(2019)\citenamefont {Levine}, \citenamefont {Turner}, \citenamefont {Kehayias}, \citenamefont {Hart}, \citenamefont {Langellier}, \citenamefont {Trubko}, \citenamefont {Glenn}, \citenamefont {Fu},\ and\ \citenamefont {Walsworth}}]{LevineTurnerKehayiasHartLangellierTrubkoGlennFuWalsworth+2019+1945+1973}%
  \BibitemOpen
  \bibfield  {author} {\bibinfo {author} {\bibfnamefont {E.~V.}\ \bibnamefont {Levine}}, \bibinfo {author} {\bibfnamefont {M.~J.}\ \bibnamefont {Turner}}, \bibinfo {author} {\bibfnamefont {P.}~\bibnamefont {Kehayias}}, \bibinfo {author} {\bibfnamefont {C.~A.}\ \bibnamefont {Hart}}, \bibinfo {author} {\bibfnamefont {N.}~\bibnamefont {Langellier}}, \bibinfo {author} {\bibfnamefont {R.}~\bibnamefont {Trubko}}, \bibinfo {author} {\bibfnamefont {D.~R.}\ \bibnamefont {Glenn}}, \bibinfo {author} {\bibfnamefont {R.~R.}\ \bibnamefont {Fu}}, \ and\ \bibinfo {author} {\bibfnamefont {R.~L.}\ \bibnamefont {Walsworth}},\ }\href {\doibase doi:10.1515/nanoph-2019-0209} {\bibfield  {journal} {\bibinfo  {journal} {Nanophotonics}\ }\textbf {\bibinfo {volume} {8}},\ \bibinfo {pages} {1945} (\bibinfo {year} {2019})}\BibitemShut {NoStop}%
\bibitem [{\citenamefont {Schoenfeld}\ and\ \citenamefont {Harneit}(2011)}]{PhysRevLett.106.030802}%
  \BibitemOpen
  \bibfield  {author} {\bibinfo {author} {\bibfnamefont {R.~S.}\ \bibnamefont {Schoenfeld}}\ and\ \bibinfo {author} {\bibfnamefont {W.}~\bibnamefont {Harneit}},\ }\href {\doibase 10.1103/PhysRevLett.106.030802} {\bibfield  {journal} {\bibinfo  {journal} {Phys. Rev. Lett.}\ }\textbf {\bibinfo {volume} {106}},\ \bibinfo {pages} {030802} (\bibinfo {year} {2011})}\BibitemShut {NoStop}%
\bibitem [{\citenamefont {Wang}\ \emph {et~al.}(2021)\citenamefont {Wang}, \citenamefont {McPherson}, \citenamefont {Kadado}, \citenamefont {Brandt}, \citenamefont {Edwards}, \citenamefont {Casey},\ and\ \citenamefont {Curro}}]{PhysRevApplied.16.054014}%
  \BibitemOpen
  \bibfield  {author} {\bibinfo {author} {\bibfnamefont {Z.}~\bibnamefont {Wang}}, \bibinfo {author} {\bibfnamefont {C.}~\bibnamefont {McPherson}}, \bibinfo {author} {\bibfnamefont {R.}~\bibnamefont {Kadado}}, \bibinfo {author} {\bibfnamefont {N.}~\bibnamefont {Brandt}}, \bibinfo {author} {\bibfnamefont {S.}~\bibnamefont {Edwards}}, \bibinfo {author} {\bibfnamefont {W.}~\bibnamefont {Casey}}, \ and\ \bibinfo {author} {\bibfnamefont {N.}~\bibnamefont {Curro}},\ }\href {\doibase 10.1103/PhysRevApplied.16.054014} {\bibfield  {journal} {\bibinfo  {journal} {Phys. Rev. Appl.}\ }\textbf {\bibinfo {volume} {16}},\ \bibinfo {pages} {054014} (\bibinfo {year} {2021})}\BibitemShut {NoStop}%
\bibitem [{\citenamefont {Bucher}\ \emph {et~al.}(2019)\citenamefont {Bucher}, \citenamefont {Aude~Craik}, \citenamefont {Backlund}, \citenamefont {Turner}, \citenamefont {Ben~Dor}, \citenamefont {Glenn},\ and\ \citenamefont {Walsworth}}]{Bucher2019}%
  \BibitemOpen
  \bibfield  {author} {\bibinfo {author} {\bibfnamefont {D.~B.}\ \bibnamefont {Bucher}}, \bibinfo {author} {\bibfnamefont {D.~P.~L.}\ \bibnamefont {Aude~Craik}}, \bibinfo {author} {\bibfnamefont {M.~P.}\ \bibnamefont {Backlund}}, \bibinfo {author} {\bibfnamefont {M.~J.}\ \bibnamefont {Turner}}, \bibinfo {author} {\bibfnamefont {O.}~\bibnamefont {Ben~Dor}}, \bibinfo {author} {\bibfnamefont {D.~R.}\ \bibnamefont {Glenn}}, \ and\ \bibinfo {author} {\bibfnamefont {R.~L.}\ \bibnamefont {Walsworth}},\ }\href {\doibase 10.1038/s41596-019-0201-3} {\bibfield  {journal} {\bibinfo  {journal} {Nature Protocols}\ }\textbf {\bibinfo {volume} {14}},\ \bibinfo {pages} {2707} (\bibinfo {year} {2019})}\BibitemShut {NoStop}%
\bibitem [{\citenamefont {Roy}\ \emph {et~al.}(2011)\citenamefont {Roy}, \citenamefont {Bandyopadhyay},\ and\ \citenamefont {Atulasimha}}]{10.1063/1.3624900}%
  \BibitemOpen
  \bibfield  {author} {\bibinfo {author} {\bibfnamefont {K.}~\bibnamefont {Roy}}, \bibinfo {author} {\bibfnamefont {S.}~\bibnamefont {Bandyopadhyay}}, \ and\ \bibinfo {author} {\bibfnamefont {J.}~\bibnamefont {Atulasimha}},\ }\href {\doibase 10.1063/1.3624900} {\bibfield  {journal} {\bibinfo  {journal} {Applied Physics Letters}\ }\textbf {\bibinfo {volume} {99}},\ \bibinfo {pages} {063108} (\bibinfo {year} {2011})},\ \Eprint {http://arxiv.org/abs/https://pubs.aip.org/aip/apl/article-pdf/doi/10.1063/1.3624900/14456471/063108\_1\_online.pdf} {https://pubs.aip.org/aip/apl/article-pdf/doi/10.1063/1.3624900/14456471/063108\_1\_online.pdf} \BibitemShut {NoStop}%
\bibitem [{\citenamefont {Degen}\ \emph {et~al.}(2017)\citenamefont {Degen}, \citenamefont {Reinhard},\ and\ \citenamefont {Cappellaro}}]{RevModPhys.89.035002}%
  \BibitemOpen
  \bibfield  {author} {\bibinfo {author} {\bibfnamefont {C.~L.}\ \bibnamefont {Degen}}, \bibinfo {author} {\bibfnamefont {F.}~\bibnamefont {Reinhard}}, \ and\ \bibinfo {author} {\bibfnamefont {P.}~\bibnamefont {Cappellaro}},\ }\href {\doibase 10.1103/RevModPhys.89.035002} {\bibfield  {journal} {\bibinfo  {journal} {Rev. Mod. Phys.}\ }\textbf {\bibinfo {volume} {89}},\ \bibinfo {pages} {035002} (\bibinfo {year} {2017})}\BibitemShut {NoStop}%
\bibitem [{\citenamefont {Hart}\ \emph {et~al.}(2021)\citenamefont {Hart}, \citenamefont {Schloss}, \citenamefont {Turner}, \citenamefont {Scheidegger}, \citenamefont {Bauch},\ and\ \citenamefont {Walsworth}}]{PhysRevApplied.15.044020}%
  \BibitemOpen
  \bibfield  {author} {\bibinfo {author} {\bibfnamefont {C.~A.}\ \bibnamefont {Hart}}, \bibinfo {author} {\bibfnamefont {J.~M.}\ \bibnamefont {Schloss}}, \bibinfo {author} {\bibfnamefont {M.~J.}\ \bibnamefont {Turner}}, \bibinfo {author} {\bibfnamefont {P.~J.}\ \bibnamefont {Scheidegger}}, \bibinfo {author} {\bibfnamefont {E.}~\bibnamefont {Bauch}}, \ and\ \bibinfo {author} {\bibfnamefont {R.~L.}\ \bibnamefont {Walsworth}},\ }\href {\doibase 10.1103/PhysRevApplied.15.044020} {\bibfield  {journal} {\bibinfo  {journal} {Phys. Rev. Appl.}\ }\textbf {\bibinfo {volume} {15}},\ \bibinfo {pages} {044020} (\bibinfo {year} {2021})}\BibitemShut {NoStop}%
\bibitem [{\citenamefont {Mizuno}\ \emph {et~al.}(2020)\citenamefont {Mizuno}, \citenamefont {Ishiwata}, \citenamefont {Masuyama}, \citenamefont {Iwasaki},\ and\ \citenamefont {Hatano}}]{Mizuno2020}%
  \BibitemOpen
  \bibfield  {author} {\bibinfo {author} {\bibfnamefont {K.}~\bibnamefont {Mizuno}}, \bibinfo {author} {\bibfnamefont {H.}~\bibnamefont {Ishiwata}}, \bibinfo {author} {\bibfnamefont {Y.}~\bibnamefont {Masuyama}}, \bibinfo {author} {\bibfnamefont {T.}~\bibnamefont {Iwasaki}}, \ and\ \bibinfo {author} {\bibfnamefont {M.}~\bibnamefont {Hatano}},\ }\href {\doibase 10.1038/s41598-020-68404-5} {\bibfield  {journal} {\bibinfo  {journal} {Scientific Reports}\ }\textbf {\bibinfo {volume} {10}},\ \bibinfo {pages} {11611} (\bibinfo {year} {2020})}\BibitemShut {NoStop}%
\bibitem [{\citenamefont {Yang}\ \emph {et~al.}(2022)\citenamefont {Yang}, \citenamefont {Vallabhapurapu}, \citenamefont {Sewani}, \citenamefont {Isarov}, \citenamefont {Firgau}, \citenamefont {Adambukulam}, \citenamefont {Johnson}, \citenamefont {Pla},\ and\ \citenamefont {Laucht}}]{10.1119/5.0075519}%
  \BibitemOpen
  \bibfield  {author} {\bibinfo {author} {\bibfnamefont {Y.}~\bibnamefont {Yang}}, \bibinfo {author} {\bibfnamefont {H.~H.}\ \bibnamefont {Vallabhapurapu}}, \bibinfo {author} {\bibfnamefont {V.~K.}\ \bibnamefont {Sewani}}, \bibinfo {author} {\bibfnamefont {M.}~\bibnamefont {Isarov}}, \bibinfo {author} {\bibfnamefont {H.~R.}\ \bibnamefont {Firgau}}, \bibinfo {author} {\bibfnamefont {C.}~\bibnamefont {Adambukulam}}, \bibinfo {author} {\bibfnamefont {B.~C.}\ \bibnamefont {Johnson}}, \bibinfo {author} {\bibfnamefont {J.~J.}\ \bibnamefont {Pla}}, \ and\ \bibinfo {author} {\bibfnamefont {A.}~\bibnamefont {Laucht}},\ }\href {\doibase 10.1119/5.0075519} {\bibfield  {journal} {\bibinfo  {journal} {American Journal of Physics}\ }\textbf {\bibinfo {volume} {90}},\ \bibinfo {pages} {550} (\bibinfo {year} {2022})},\ \Eprint {http://arxiv.org/abs/https://pubs.aip.org/aapt/ajp/article-pdf/90/7/550/19799837/550\_1\_online.pdf} {https://pubs.aip.org/aapt/ajp/article-pdf/90/7/550/19799837/550\_1\_online.pdf} \BibitemShut
  {NoStop}%
\bibitem [{\citenamefont {Tetienne}\ \emph {et~al.}(2018)\citenamefont {Tetienne}, \citenamefont {Broadway}, \citenamefont {Lillie}, \citenamefont {Dontschuk}, \citenamefont {Teraji}, \citenamefont {Hall}, \citenamefont {Stacey}, \citenamefont {Simpson},\ and\ \citenamefont {Hollenberg}}]{s18041290}%
  \BibitemOpen
  \bibfield  {author} {\bibinfo {author} {\bibfnamefont {J.-P.}\ \bibnamefont {Tetienne}}, \bibinfo {author} {\bibfnamefont {D.~A.}\ \bibnamefont {Broadway}}, \bibinfo {author} {\bibfnamefont {S.~E.}\ \bibnamefont {Lillie}}, \bibinfo {author} {\bibfnamefont {N.}~\bibnamefont {Dontschuk}}, \bibinfo {author} {\bibfnamefont {T.}~\bibnamefont {Teraji}}, \bibinfo {author} {\bibfnamefont {L.~T.}\ \bibnamefont {Hall}}, \bibinfo {author} {\bibfnamefont {A.}~\bibnamefont {Stacey}}, \bibinfo {author} {\bibfnamefont {D.~A.}\ \bibnamefont {Simpson}}, \ and\ \bibinfo {author} {\bibfnamefont {L.~C.~L.}\ \bibnamefont {Hollenberg}},\ }\href {\doibase 10.3390/s18041290} {\bibfield  {journal} {\bibinfo  {journal} {Sensors}\ }\textbf {\bibinfo {volume} {18}} (\bibinfo {year} {2018}),\ 10.3390/s18041290}\BibitemShut {NoStop}%
\bibitem [{\citenamefont {Lang}\ \emph {et~al.}(2019)\citenamefont {Lang}, \citenamefont {Madhavan}, \citenamefont {Tetienne}, \citenamefont {Broadway}, \citenamefont {Hall}, \citenamefont {Teraji}, \citenamefont {Monteiro}, \citenamefont {Stacey},\ and\ \citenamefont {Hollenberg}}]{PhysRevA.99.012110}%
  \BibitemOpen
  \bibfield  {author} {\bibinfo {author} {\bibfnamefont {J.~E.}\ \bibnamefont {Lang}}, \bibinfo {author} {\bibfnamefont {T.}~\bibnamefont {Madhavan}}, \bibinfo {author} {\bibfnamefont {J.-P.}\ \bibnamefont {Tetienne}}, \bibinfo {author} {\bibfnamefont {D.~A.}\ \bibnamefont {Broadway}}, \bibinfo {author} {\bibfnamefont {L.~T.}\ \bibnamefont {Hall}}, \bibinfo {author} {\bibfnamefont {T.}~\bibnamefont {Teraji}}, \bibinfo {author} {\bibfnamefont {T.~S.}\ \bibnamefont {Monteiro}}, \bibinfo {author} {\bibfnamefont {A.}~\bibnamefont {Stacey}}, \ and\ \bibinfo {author} {\bibfnamefont {L.~C.~L.}\ \bibnamefont {Hollenberg}},\ }\href {\doibase 10.1103/PhysRevA.99.012110} {\bibfield  {journal} {\bibinfo  {journal} {Phys. Rev. A}\ }\textbf {\bibinfo {volume} {99}},\ \bibinfo {pages} {012110} (\bibinfo {year} {2019})}\BibitemShut {NoStop}%
\bibitem [{\citenamefont {Caciagli}\ \emph {et~al.}(2018)\citenamefont {Caciagli}, \citenamefont {Baars}, \citenamefont {Philipse},\ and\ \citenamefont {Kuipers}}]{CACIAGLI2018423}%
  \BibitemOpen
  \bibfield  {author} {\bibinfo {author} {\bibfnamefont {A.}~\bibnamefont {Caciagli}}, \bibinfo {author} {\bibfnamefont {R.~J.}\ \bibnamefont {Baars}}, \bibinfo {author} {\bibfnamefont {A.~P.}\ \bibnamefont {Philipse}}, \ and\ \bibinfo {author} {\bibfnamefont {B.~W.}\ \bibnamefont {Kuipers}},\ }\href {\doibase https://doi.org/10.1016/j.jmmm.2018.02.003} {\bibfield  {journal} {\bibinfo  {journal} {Journal of Magnetism and Magnetic Materials}\ }\textbf {\bibinfo {volume} {456}},\ \bibinfo {pages} {423} (\bibinfo {year} {2018})}\BibitemShut {NoStop}%
\bibitem [{\citenamefont {Schloss}\ \emph {et~al.}(2018)\citenamefont {Schloss}, \citenamefont {Barry}, \citenamefont {Turner},\ and\ \citenamefont {Walsworth}}]{PhysRevApplied.10.034044}%
  \BibitemOpen
  \bibfield  {author} {\bibinfo {author} {\bibfnamefont {J.~M.}\ \bibnamefont {Schloss}}, \bibinfo {author} {\bibfnamefont {J.~F.}\ \bibnamefont {Barry}}, \bibinfo {author} {\bibfnamefont {M.~J.}\ \bibnamefont {Turner}}, \ and\ \bibinfo {author} {\bibfnamefont {R.~L.}\ \bibnamefont {Walsworth}},\ }\href {\doibase 10.1103/PhysRevApplied.10.034044} {\bibfield  {journal} {\bibinfo  {journal} {Phys. Rev. Appl.}\ }\textbf {\bibinfo {volume} {10}},\ \bibinfo {pages} {034044} (\bibinfo {year} {2018})}\BibitemShut {NoStop}%
\end{thebibliography}
\end{document}